\documentclass[12pt,preprintnumbers]{revtex4}

\input epsf
\usepackage{amsmath}
\usepackage{amsfonts}
\usepackage{amssymb}%
\usepackage{graphicx}
\usepackage{epsfig}

\newcommand{\bel}[1]{\begin{equation}\label{#1}}
\newcommand{\ee}{\end{equation}}

\newcommand{\reffig}[1]{Fig.~\ref{fig:#1}}

\newcommand{\beq}{\begin{eqnarray}}% can be used as {equation} or {eqnarray}
\newcommand{\eeq}{\end{eqnarray}}

\def\mn{{\mu\nu}}

\def\be{\begin{equation}}
\def\ee{\end{equation}}
\newcommand{\eref}[1]{(\ref{#1})}
\newcommand{\Eref}[1]{Eq.~(\ref{#1})}
\newcommand{\rem}[1]{}

\def\OO{{\cal O}}
\def\ZZ{{\bf Z}}

\def\shat{\hat s}
\def\gsim{\agt}
\def\lsim{\alt}
\def\fourgamma{\gamma\gamma\gamma\gamma}
\def\Imm{{\rm Im}}

\begin{document}

\preprint{RUNHETC-2009-31}

\title{A Simple-Minded Unitarity Constraint\\ and
an Application to Unparticles.}
  
\author{Antonio Delgado}
\affiliation{Department of Physics, 225 Nieuwland Science Hall,
Notre Dame, IN 46556}
\author{Matthew J. Strassler}
\affiliation{Department of Physics and Astronomy,
Rutgers University, Piscataway, NJ 08854}

\begin{abstract}
Unitarity, a powerful constraint on new physics,
has not always been properly accounted for
in the context of hidden sectors.
%The cross-section of any process involving a hidden sector is
%obviously less than the total cross-section of all such processes.
%The latter can be computed, using unitarity, if the standard model is
%coupled via a pointlike interaction to a scale-invariant hidden sector
%(the ``unparticle'' scenario.)  This constrains any individual
%process, at all energy scales.  
Feng, Rajaraman and Tu have suggested that large (pb to nb)
multi-photon or multi-lepton sugnals could be generated at the LHC
through the three-point functions of a conformally-invariant hidden
sector (an ``unparticle'' sector.)  Because of the conformal
invariance, the kinematic distributions are calculable. However, the
cross-sections for many such processes grow rapidly with energy, and
at some high scale, to preserve unitarity, conformal invariance must
break down.  Requiring that conformal invariance not be broken, and
that no signals be already observed at the Tevatron, we obtain a
strong unitarity bound on multi-photon events at the (10 TeV) LHC.
For the model of Feng et al., even with extremely
conservative assumptions, cross-sections must be below 25 fb, and
for operator dimension near 2, well below 1 fb.  In more general
models, four-photon signals could still reach cross-sections of a few
pb, though bounds below 200 fb are more typical.  Our methods apply to
a wide variety of other processes and settings.

\end{abstract}

\maketitle

\newpage

\section{Introduction}

The current era is dominated by hadron colliders, where small signals
must be extracted from very large data sets.  In order that new
physics of an unfamiliar sort not be missed, it is important to
consider a wide variety of possible signals that the experimenters
might encounter.  In this spirit, there has been considerable activity
aimed at thinking broadly about reasonable non-minimal extensions of
the standard model Higgs sector, of minimal supersymmetric models, and
so forth.  While there are strong motivations for each of these
classes of models, the simplicity of their minimal versions is
motivated mainly by aesthetic considersations.  Moreover, the extra
particles in non-minimal versons can lead to completely different
phenomenological signals from those arising in the minimal versions.
Given the baroque nature of the standard model, we would be unwise
when addressing important issues in particle physics not to consider
the possibility of particles and forces beyond the minimal set
required.

Considerable attention has been paid recently to hidden sectors that
couple to the standard model at or near the TeV scale.  These include
``hidden valleys'' \cite{hv1,hv2,hvsusy}, new sectors with mass gaps
and non-trivial dynamics, which lead to new light neutral particles,
often produced in clusters and with a boost, and possibly with
macroscopically long lifetimes.  Hidden valleys are especially natural
hosts for dark matter, and indeed a class of hidden valley models
\cite{hvdark} are a popular explanation for current anomalies in
dark-matter experiments.

Work on hidden sectors also includes a great deal of research on
conformally invariant hidden sectors, dubbed ``unparticles'' in
\cite{Un1,Un2, allunparticle}
(see also \cite{RS2,HEIDI}).  New sectors with
conformally-invariant physics (or at least scale-invariant physics,
though there are no known examples of theories in four dimensions with
scale invariance but without conformal invariance) can produce large
missing-tranverse-momentum (``MET'') signals, and can produce smaller,
but potentially still dramatic, visible effects.  However, the
literature on this subject is full of contradictions, and many claims
of interesting effects have been criticized.  This has left the
experimental community without clear guidance as to how to search for
hidden sectors of this type.

Our goal in this paper is to bring some clarity, through simple
arguments, to a claim \cite{FengRajTu} that large production rates for
multi-particle final states can be generated through the three-point
function of hidden sector operators that couple to the standard model.
(Other work emphasizing the importance of higher-point functions,
often called ``unparticle interactions,'' can be found in
\cite{hvun,GeorgiUnInts}.  Additional subtle issues are addressed in
\cite{Un2,allunparticle,unFox,unDelgado,Grinstein}.) We consider
specifically the mechanism discussed by Feng, Rajaraman and Tu in
\cite{FengRajTu}, slightly generalized.  In \cite{FengRajTu} it was
pointed out that (for example) if a scalar primary operator $\OO$ in
the hidden sector couples to two gluons and also to two photons, and
has a non-trivial three-point function $\langle \OO\OO\OO \rangle$,
then the process $gg\to\fourgamma$ can be generated.  Because the form
of a three-point function $\langle \OO_1\OO_2\OO_3 \rangle$ of primary
scalar operators is precisely determined in conformal field theory in
terms of the dimensions $\Delta_i$ of the three operators $\OO_i$, the
kinematics of any process of this type is precisely known.  (This is
also true in some cases for three-point functions involving operators
with non-zero spin.)  In the case considered by \cite{FengRajTu}, all
kinematic distributions can be calculated in terms of the dimension
and spin of $\OO$.

Moreover, there is only one unknown parameter, the overall coefficient
of the three-point function (equivalently the OPE coefficient
connecting $\OO\OO\to\OO$.)  In \cite{FengRajTu} it was pointed out
that as of yet there is no known bound in four-dimensions on the size
of this coefficient, and so it was suggested it could be arbitrarily
large.  Based on the limits 
%
% CITE???
%
from Fermilab on multi-photon events, it
was claimed in \cite{FengRajTu} that LHC production rates (at 14 TeV)
were little constrained, and could range as large as 4 pb for
$\Delta_\OO\sim 1.1$ and 8 nb (ten times larger than the $t\bar t$
cross-section) for $\Delta_\OO\sim 1.9$.  Given that four-photon
backgrounds are tiny, and that the photons produced in this process
have very high $p_T$, this
would be a truly spectacular signal by any measure.

In this paper we throw some amount of cold water on this possibility.
We first observe a simple-minded (and model-independent) unitarity
constraint on any hidden sector, conformal or not.  Then we show how
this specifically constrains conformally-invariant sectors, where
explicit computations are possible due to the conformal invariance.
After putting some experimental and theoretical limits on the size of
the coupling between the two sectors, we apply this constraint
specifically to the process $pp\to \gamma\gamma\gamma\gamma$.  For the
specific case studied in \cite{FengRajTu}, we find the maximum
cross-section (for LHC at 10 TeV) is actually of order 20 fb.  When we
generalize the scenario considered in \cite{FengRajTu} by allowing the
two gluons to couple to one operator $\OO_1$ and the two
photons to couple a different operator $\OO_2$, we find that the maximum
cross-section is anywhere from several pb, in the region $\Delta_1\sim 1.4$
and $\Delta_2\sim 1.1$, down to 30 fb or below for
$\Delta_1+2\Delta_2>5$.

Our methods can be applied more widely to various other processes.
They will strongly constrain four-lepton production through vector
unparticles, for example, and any other similar process.

As this paper neared completion some additional work on this subject
appeared in \cite{fourleptons,new4gamma}.  We believe that application of our
methods would affect the conclusions of these papers.  Also, in 
\cite{new4gamma} production of multiple particles through exchange of two
unparticles was considered.  While we do not address this issue in our
current paper, there are additional and related unitarity bounds on this process
which were not considered in \cite{new4gamma}.  It should also be noted
that the authors of 
\cite{new4gamma} assumed in their calculation that there is no important 
four-point function among the hidden-sector operators, 
which is not universally true.

The paper is organized as follows.  We will explain our unitary bound
in section II. After some general comments in section III about
applications to unparticle sectors, we will show how to apply it to
the specific case of $gg\to\fourgamma$ in section IV.  Section V will
be devoted to obtaining a bound on the scale $\Lambda_1$
characterizing the coupling between the two gluons 
and the unparticle sector. In section VI we will
calculate the numerical bounds on $pp\to \gamma \gamma\gamma\gamma$.  We will
comment on other possible processes in section VII, and state some
conclusions in section VIII.

\section{A trivial unitarity bound}

We begin by pointing out an essentially trivial but rigorous unitarity
bound that governs parton-parton cross-sections for hidden-sector
production.  The point, simply stated, is that {\it no one process
that involves the hidden sector can have a rate that exceeds the total
rate for all such processes.}

This simple-minded and obvious point becomes useful when the total
rate can be computed.  Among the situations where
this is possible is the case when the hidden sector is a conformal
field theory to which the standard model (SM) couples via a local
interaction.  In this case the total cross-section is given by the
square of a standard-model amplitude times the imaginary part of a
two-point function of a local operator in the conformal field theory
(recently given the name ``unparticle propagator'' \cite{Un1}.)  
Consequently, one may calculate the bound on the sum of all processes
involving the hidden sector.

Let us make a technically more precise statement of this unitarity
bound.  Suppose the interaction between the two sectors is governed by
a local interaction, for example of the form
\begin{equation}\label{coupling1}
{1\over \Lambda^\delta}\psi_A\psi_B \OO
\end{equation}
where $\psi_{A,B}$ are SM fields that create the SM partons $A,B$,
and $\OO$ is a gauge-invariant operator in the hidden sector that
carries no SM charges.  (We take $\OO$ to be spinless for
the moment, but our statements generalize for any spin.)  

We consider first a process $A B \to X$ where $X$ is a state in the
hidden-sector Hilbert space.  We will refer to the sum over all such
states as $A B \to \left\{X\right\}$. Then the optical theorem assures
that for center-of-mass momentum $q^\mu=q_A^\mu+q_B^\mu$ and
center-of-mass energy $\sqrt{\shat}=q^2$,
\beq\label{optibound}
\sigma(AB\to \{X\}; \shat) &\equiv& 
\sum_X
\sigma(AB\to X; \shat)\cr
& \ & \cr
&=& \frac{\Imm(AB\to \{X\}\to AB)}{ \shat}
=
\frac{\left|\langle AB|\psi_B\psi_A|0\rangle\right|^2}{ \Lambda^{2\delta}}
\frac{\Imm \left[i\langle 0|\OO(q) \OO(-q)|0\rangle\right]}{\shat} \ .
\cr
& \ &
\eeq
%CHECK APPEARANCE???
%
Corrections to this last formula are smaller than the leading
expression by a factor of order
$(\shat/\Lambda^2)^\delta$.  We simplify notation by defining
\begin{equation}\label{newdefs}
f_{AB}\equiv \langle AB|\psi_B\psi_A|0\rangle \ \ ; \ \
G_{\OO}(q;\Lambda) \equiv  i\langle 0|\OO(q) \OO(-q)|0\rangle
\end{equation}
so that 
\begin{equation}\label{optibound2}
\sigma(AB\to \{X\}; \shat) 
=
 {1\over\Lambda^{2\delta}\shat}|f_{AB}|^2 \Imm\left[G_\OO(q;\Lambda)\right]
\end{equation}
with $\shat=q^2$.

 We are effectively assuming that the two sectors are weakly coupled
to one another, so that the Hilbert space factors into a SM part and a
hidden-sector part.  This is true in the limit $\Lambda\to\infty$, and
the corrections to this assumption should be small as long as momenta
are small compared, naively, to $4 \pi\Lambda$.  Actually, whether the
condition involves $4\pi\Lambda$ or a somewhat smaller scale depends,
as we will see, on the operator and on $A,B$.  Also we have assumed
here that any process generated by two separate couplings of the
initial state to the hidden sector, such as considered in
\cite{new4gamma}, is subleading compared to the effect of a single
such coupling.  If this is not the case, self-consistency problems
arise, which we will not address here.

\begin{figure}[htbp]  
\begin{center}  
\leavevmode
\vskip 0.3in
\includegraphics[width=0.65\textwidth]{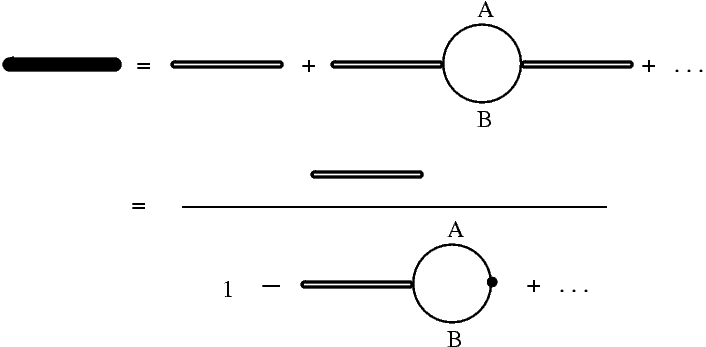} 
\end{center}
\vskip -0.00in 
 \caption{The full two point function for $\OO$ (filled line) differs
from the conformal two-point function (unfilled line) by loops of standard
model particles; these can be resummed as usual into a geometric series.}
\label{fig:GSig}
\end{figure}

Importantly, as emphasized by our notation, the two-point function of
$\OO$ that appears here is the {\it complete} two-point function,
which includes all effects that depend on $\Lambda$ from the
interaction \eref{coupling1}, along with any other interactions
between the SM and hidden sectors.  Let us define the
two-point function of $\OO$ in the limit $\Lambda\to\infty$ to be
\begin{equation}\label{OOhidonly}
G_\OO^{(0)}(q)\equiv\lim_{\Lambda\to\infty}G_\OO(q;\Lambda)
\end{equation}
The difference between this function and the full two-point function 
includes terms
such as
\begin{equation}\label{corrections2OO}
G_\OO(q;\Lambda)= G_\OO^{(0)}(q) +i G_\OO^{(0)}(q)^2
\frac{1}{\Lambda^{2\delta}}\int \frac{d^4k}{(2\pi)^4} 
<0|\psi_B(k)\psi_A(q-k) \psi_B(-k)\psi_A(k-q)|0>+\dots
\end{equation}
as shown in \reffig{GSig}.
This particular type of correction sums as usual into a geometric series 
\begin{equation}\label{resummed2OO}
G_\OO(q;\Lambda)=\frac{ G_\OO^{(0)}(q)}{1- G_\OO^{(0)}(q)\Sigma(q)-\dots}
\end{equation}
where
\begin{equation}\label{Sigmadef}
\Sigma(q)={i\over \Lambda^{2\delta}}\int \frac{d^4k}{(2\pi)^4} 
<0|\psi_B(k)\psi_A(q-k) \psi_B(-k)\psi_A(k-q)|0>+\dots
\end{equation}
as in \reffig{GSig}. Other processes that connect the two sectors will
also contribute to the full two-point function.  

Suppose we demand that the full two-point function $G_\OO(q;\Lambda)$
does not differ much from its $\Lambda\to\infty$ limit
$G_\OO^{(0)}(q)$ --- that is, that the interaction with the SM sector
does not strongly alter the hidden sector in the energy regime of
interest.  (In particular, if the hidden sector is conformal in the
$\Lambda\to\infty$ limit, then we are demanding that it remain so to a
good approximation.)  Then any process such as $AB\to P_1 P_2\dots +
X_0$, where $P_i$ are SM particles and $X_0$ is any hidden sector
state, and where $P_i$ are produced dominantly through SM-hidden
sector interactions suppressed by $1/\Lambda$ to some power, can be
bounded.  In particular, this process will appear in the imaginary
part of the full two-point function, suppressed by powers of
$1/\Lambda$ to some power.  The statement that the $1/\Lambda$
corrections to $G_\OO(q,\Lambda)$ are small, applied to its imaginary
part, then implies that
\beq\label{boundonABCCCX}
\sum_{\{X_0\}}\sigma(AB\to P_1 P_2\dots + X_0)&<& 
{1\over\Lambda^{2\delta}\shat}|f_{AB}|^2
\Imm\left[ G_\OO(q;\Lambda)-
         G_\OO^{(0)}(q) \right]\ \cr
& \ & \ \cr
& \ll &
{1\over\Lambda^{2\delta}\shat}|f_{AB}|^2
\Imm\left[ G_\OO(q;\Lambda)\right] \ \cr & \ & \ \cr
&\approx& 
{1\over\Lambda^{2\delta}\shat}|f_{AB}|^2
\Imm\left[  G_\OO^{(0)}(q)\right]
 \approx \sigma(AB\to \{X\})
\eeq
The sum over $X_0$ is over any subset of (and including possibly all)
allowed hidden-sector states.  The corrections to the last approximate
equality vanish as $\Lambda\to\infty$.  Note the expressions in the
first line are of higher order in $1/\Lambda$ than those in the last
line, since by definition $G_\OO(q;\Lambda)\to G_\OO^{(0)}$ as
$\Lambda\to\infty$.  Therefore this is obviously true when $\Lambda\gg
q$.  But for LHC signals we will be interested in the consequences
when $q$ and $\Lambda$ are not well separated.

In English, the relations \eref{boundonABCCCX} state the following.
The first inequality says that the process in question is found in the
imaginary part of $G_\OO$ {\it which does not appear in} $G_\OO^{(0)}$,
since the latter contains only processes involving the hidden sector
alone.  The second inequality says that the difference between $G_\OO$
and $G_\OO^{(0)}$ cannot be large, if conformal symmetry is valid.
The third approximate equality restates that $G_\OO$ and $G_\OO^{(0)}$
must be similar, so we may use either one.  The final approximate
equality comes from Eq.~\eref{optibound2}.  The last two inequalities
become equalities in the limit $\Lambda\to\infty$.  

It is crucial that the constraint \eref{boundonABCCCX} depends on $q$,
or $\sqrt{\shat}$, the partonic collision energy, not directly on the
collider energy $\sqrt s$.  Thus, {\it at a hadron collider, this constraint
must be applied at all relevant values of $\sqrt{ \shat}$.}

\section{Application to Conformal Hidden Sectors (Unparticles)}

\subsection{Conformal invariance must break down}

If the hidden sector is conformal, then $ G_\OO^{(0)}(q)$ is
determined, up to a normalization constant.  The canonical
normalization is taken so that in position space the time-ordered
two-point function is $1/(4 \pi^2 x^2)^{\Delta}$ (up to contact terms
at $x=0$); any other normalization factor can be absorbed into
$\Lambda$.  The Fourier transform to momentum space yields
\begin{equation}\label{G0normed}
 G_\OO^{(0)}(q)=  \frac{1}{(4\pi)^{2\Delta-2}}
\frac{\Gamma[2-\Delta]}{%(4\pi)^{2\Delta-2}
\Gamma[\Delta]} 
(-q^2 - i \epsilon)^{\Delta-2} \ .
\end{equation}
Our normalization is the same as that used in \cite{Un1}, simplified
by the use of Gamma-function identities.

Suppose we want to use conformal invariance to predict something in
the hidden sector.  Then we must demand that any corrections to the
two-point function are small compared to the two-point function
itself, which then implies the bound \eref{boundonABCCCX}.  In
particular, for any particular process (such as $gg\to\fourgamma$, as
we will consider below)
in which only SM particles
$P_i$ are produced through the hidden sector,
\begin{equation}\label{boundonABKKKK}
\sigma(AB\to \{X\}\to P_1 P_2\dots P_n) \ll \sigma(AB\to \{X\})
\end{equation}
In fact the bound is much stronger than this; the {\it sum} of
cross-sections for {\it all} such processes, producing any standard
model particles and hidden-sector states, is smaller than
$\sigma(AB\to \{X\})$.  If conformal invariance predicts cross-sections
that violate this condition, then it is 
conformal invariance itself that
must be violated, and thus it
cannot be used to make predictions.

To illustrate the issues, let us consider a Lagrangian with three
terms that couple the SM to the hidden sector through couplings to
scalar hidden-sector operators, of the form
\begin{equation}\label{Lagrangian}
\delta{\cal L} = {1\over \Lambda_1^{\delta_1}}\OO_1 \psi_A\psi_B + 
{1\over \Lambda_2^{\delta_2}}\OO_2 \psi_1\psi_2 +
{1\over \Lambda_3^{\delta_3}}\OO_3 \psi_3\psi_4 
\end{equation}
Here $\delta_1=\Delta_1+{\rm dim}\psi_A + {\rm dim}\psi_B-4$, and
similarly for $\delta_2,\delta_3$.  (For the moment we take all three
operators $\OO_i$ to be distinct; the case where the operators are
related will be dealt with later.  We also assume $\delta_i>0$; we
will discuss this assumption later.  The standard model fields
$\psi_i$, which create particles $P_i$, may or may not be different
from one another; we make no assumptions about them as yet.) Then,
purely from dimensional analysis, we have
\begin{equation}\label{sigmatot}
\sigma(AB\to \{X\}; \shat) = \frac
{N_0(\Delta_1)}{\shat}
\left(\frac{\sqrt{\shat}}{\Lambda}\right)^{2\delta_1}
\end{equation}
where $N_0$ is a constant calculable from conformal invariance alone and
which depends only on $\Delta_1$ and on $|f_{AB}|^2$.
Meanwhile,
\begin{equation}\label{ABKKKKsigma}
\sigma(AB\to P_i; \shat) = \frac{|C_{123}|^2}{\shat}
N_{P_i}(\Delta_1,\Delta_2,\Delta_3)
\left(\frac{\sqrt{\shat}}
{\Lambda_1}\right)^{2\delta_1}
\left(\frac{\sqrt{\shat}}{\Lambda_2}\right)^{2\delta_2}
\left(\frac{\sqrt{\shat}}{\Lambda_3}\right)^{2\delta_3}
\end{equation}
Here, as emphasized by \cite{FengRajTu}, $N_{P_i}$ is a constant which
is determined by the dimensions of the operators $\OO_i$.
We will see we do not need its exact form.  The OPE coefficient
$C_{123}$ for $\OO_1\OO_2\to \OO_3$ determines the normalization of
the $\langle\OO_1 \OO_2 \OO_3\rangle$ three-point function.  Again,
its value will not be needed for our discussion.
% In fact, it will be useful for us to rewrite this as
%\begin{equation}\label{ABCDEF}
%\sigma(AB\to P_i; \shat) = 
%\left[N_{P_i}(\Delta_1,\Delta_2,\Delta_3,C_{123})
%\left(\frac{1}{\Lambda_2}\right)^{2\delta_2}
%\left(\frac{1}{\Lambda_3}\right)^{2\delta_3}\right]
%\frac{\hat s^{\delta_1+\delta_2+\delta_3}}
%{\Lambda_1^{2\delta_1}}
%\end{equation}

These expressions are valid up to the scale $\shat$ where conformal
predictions break down.  {\it A sufficient condition for such a
breakdown would be that $\sigma(AB\to P_i;\shat)\sim \sigma(AB\to
\{X\}; \shat)$.}  If $\delta_2+\delta_3>0$, as we are assuming at the
moment, then $\sigma(AB\to P_i;\hat s)$ grows faster with energy than
$\sigma(AB\to \{X\}; \shat)$.  Thus there is always a scale
$\shat_{max}$ at which the expressions in Eqs.~\eref{sigmatot} and
\eref{ABKKKKsigma} become equal.  At best, conformal invariance can be
used to make predictions only up to this scale.  At scales of order or
larger than $\shat_{max}$ there must be large corrections to the
two-point function of $\OO_1$.  When this happens, we can predict
neither $\sigma(AB\to \{X\})$ --- which requires the two-point
function directly --- nor $\sigma(AB\to P_i)$ --- which is predicted
using the special form of the three-point function, whose derivation
requires that the two-point function of $\OO_1$ be its conformal form.

\subsection{Motivation for studying $gg\to\gamma\gamma\gamma\gamma$}
\label{subsec:Motivate}

We must first decide what physical processes to study, which requires
us to address some subtle points.  The reader only interested in our 
results can jump to Sec.~\ref{sec:boundon4gam}.

We will focus on processes involving gauge bosons only.  Our reasoning
is the following.  The largest effects from hidden sectors would come
from low dimension operators.  Scalar operators have the lowest
possible dimensions, as is well known from unitarity bounds
\cite{Mack}. (See also \cite{SeibergNAD,ISthreeD} for other famous and
important applications of these unitarity bounds.)  We will discuss
operators of non-zero spin in Sec.~\ref{sec:otherprocesses}.  The only
standard-model scalar operators of low dimension are of the form (1)
$F_\mn F^\mn$ or $F_\mn\tilde F^\mn$ for one of the standard model
field strengths, (2) the Higgs boson bilinear $H^\dagger H$, or (3) $f
H f'$, where $f$ is a SM fermion doublet and $f'$ is a SM fermion
singlet.

Large couplings of the form $f H f' \OO$ break chiral flavor
symmetries and are extremely dangerous, especially for the light
quarks found in the proton.  Without powerful symmetries or
fine-tuning, these interactions will generically induce large and
excluded flavor-changing neutral currents, through processes such as
$f \bar f'\to f' \bar f$, $f \to f' \gamma$, etc., mediated via
effects of the hidden sector.  Conversely, suppressing flavor-changing
neutral currents by choosing small couplings (i.e., choosing a very
large value for $\Lambda$), reduces all cross-sections involving the
hidden sector by factors of $s/\Lambda^2$ to a positive power.  We are
skeptical that there exists an elegant model-building strategy 
that would permit operators to couple to the light quarks
with $\Lambda$ of
order 1 TeV and $\Delta$ not far above 1 without risking large $K$-$\bar K$
mixing.  Conversely, as $\Delta$ approaches 2, our bounds come into force.
(Couplings of SM fermions to vector
unparticles do not break chiral symmetries and are much more
reasonable, but we are only considering scalar operators at the
moment.)
Consequently, it is far more natural that the initial state
coupling should be to gluons.

In the final state, fermionic couplings might have a role to play; for
example, flavor-changing constraints on couplings to bottom and top
quarks and to tau leptons are somewhat weaker, and one could imagine
larger couplings of the heavier fermions to a hidden sector.  We will
discuss the possibility of a such final states in
Sec.~\ref{sec:otherprocesses}.

Couplings to Higgs bosons are very interesting but are complicated by
the relatively large mass of the Higgs and by its expectation value.
Examples of these complications are described in
\cite{unFox,unDelgado}.  To avoid these complications in this paper,
we assume that the couplings $H^\dagger H \OO$ are not large, which in
turn implies that the rates for producing Higgs bosons are small.
In any case, Higgs bosons produced through a hidden-sector's three-point
functions will lead mostly to multi-jet states, which have large
backgrounds.

For these reasons, in order to keep our presentation simple, we will
focus on the process $gg\to\gamma\gamma\gamma\gamma$.  This case is
nice both because it is conceptually straightforward, is a spectacular
LHC signal, and was studied in some detail in \cite{FengRajTu}.  There
are nevertheless some fine-tuning issues with the signal, which we
discuss below.

\subsection{A comment on the naturalness and fine-tuning}

On general grounds, when a theory has a low-dimension scalar operator $\OO$,
fine-tuning is typically (but not automatically) necessary to avoid
generating the operator $\OO$ itself in the Lagrangian.  This operator
would then itself serve as a relevant perturbation of the conformal
field theory and conformal invariance would be lost at very high scales.

To avoid this, one would ask that any such operator transform under a 
global symmetry, so that its appearance in the Lagrangian is forbidden.
For example, $\OO$ might be a pseudoscalar instead of a scalar, or it
might transform with a minus sign under some other $\ZZ_2$ transformation, or
be part of a large multiplet under a continuous global symmetry, {\it etc.}
However, these solutions are not entirely satisfactory since we must
in general break this very symmetry to allow terms of the form
\eref{Lagrangian}.  We might require that the standard model
operator also transform under the global symmetry (for example if the
$\OO_i$ are pseudoscalars we can couple them to $F_\mn \tilde F^\mn$, instead
of $F_\mn F^\mn$ as was done in \cite{FengRajTu}.)   But this is not
entirely satisfactory, because a three-point function among three
scalar operators transforming under a $\ZZ_2$ symmetry must vanish, 
and more complicated symmetries which allow a three-point function
cannot generally be realized among SM operators.  For example, we 
cannot couple two gluons to an operator transforming under a $\ZZ_3$
symmetry without breaking that symmetry.

We might also appeal to supersymmetry to prevent $\OO$ from being
generated with a large coefficient.  In models where supersymmetry
breaking in the hidden sector occurs at a scale which is low compared
to the TeV scale, as can occur in models of gauge mediation where the
hidden sector learns of supersymmetry breaking only through its
coupling to the SM, supersymmetry can forbid the appearance of chiral
operators in the superpotential, and thus restrict the operators that
appear in the Lagrangian, down to a rather low scale.  In this case
conformal invariance would still be valid in the regime of interest.
But this is not automatic and at the very least involves non-trivial
model-building; see for example \cite{unNelson}.
%%%Double CHECK???

Even if we solve the problem of generating $\OO$ in the action, there
is still the operator $\OO^\dagger \OO$, which is usually a relevant
operator for $\Delta$ significantly less than $2$.  (Note this
operator as written may not itself be an operator of definite
dimension, but it can be written as a linear combination of such
operators, and one of them will generally have dimension less than 4.)
The question of whether $\OO^\dagger \OO$ is relevant, and, if so, why it is
not present with a large coefficient, is analogous to the question of
the small value of the Higgs boson mass.  In order even to have a
discussion about scalar operators with $\Delta$ well below 2, we must
assume either that this coefficient is somehow unnaturally suppressed,
or that it is protected by a very weakly broken supersymmetry in the
hidden sector, as in \cite{unNelson}.  (For interesting but not
yet sufficiently powerful results regarding $\OO^\dagger \OO$,
especially where $\OO$ has dimension less than 2, see \cite{RR}.)

This particular problem does not arise for $\Delta>2$,
where the square of the operator is generally irrelevant.  (It
has sometimes been erroneously suggested in the literature that scalar
``unparticles'' do not make sense for $\Delta\geq2$. But this is
simply a misinterpretation of standard singularities which require
standard operator renormalization.  All conformal field theories
contain such operators --- for example, the square of the stress
tensor.)  Our results can be applied to such operators, but as 
we will see, the bounds that we obtain for such operators are on the
verge of putting the signals out of reach of the LHC.
%%%; see for example \cite{PandS} chapter 12.)  ???

One may also ask about the coupling $H^\dagger H\OO$, where $H$ is the
standard model Higgs boson.  When the Higgs gets an expectation value,
this inevitably would generate a breaking of conformal invariance
\cite{unFox,unDelgado}. Again, if the conformal theory has an exact or
weakly broken global symmetry that acts on $\OO$, this operator would
be forbidden.  (Meanwhile the operator $H^\dagger H\OO^\dagger \OO$ is
generally irrelevant.)  In the models we consider below, any such
symmetry is broken by the couplings to the standard model.  But as long
as the high-energy physics that generates these couplings does not
directly couple the Higgs boson to the hidden sector, and a symmetry
forbids $H^\dagger H \OO$ from arising well above the TeV scale, then any
$H^\dagger H\OO$ term will be suppressed by an extra SM loop factor 
compared to the leading couplings between the two sectors, and will be
sufficiently small not to undermine our assumptions.
%Consider this further???

Thus to obtain $gg\to\fourgamma$ from a conformally invariant sector
requires quite a bit of work.  But we will finesse all these issues,
without further comment, in this paper.  This is in order to address
the specific phenomenological claims of \cite{FengRajTu}, which assume
implicitly that all these issue are resolved, but do not depend on the
precise resolution.  Also, although they are most easily explained in
the case of scalar operators, {\it our methods apply for any spin}.
At the end of this paper will briefly discuss more realistic settings,
such as a three point function involving a vector operator ${\cal
V_\mu}$, a scalar operator $\OO$, and its conjugate $\OO^\dagger$.  In
this case the operator $\OO$ could be a pseudoscalar, for instance, or
carry some additional quantum numbers,
and many of these problems would not arise.  We emphasize, therefore,
that our results are very general and would apply with similar impact in
many situations where there are no fine-tuning issues.

%However, an additional subtlety, also neglected widely in the
%literature, is the following.  As we just noted, to avoid mixing with
%$H^\dagger H$, and worse, to avoid the relevant operator $\OO$
%itself being generated in the action, and/or to avoid it
%developing an expectation value, one might wish to find a global symmetry that
%forbids linear terms in $\OO$.  But this very symmetry will forbid many
%three-point functions involving $\OO$!  

%\subsection{Naturalness and Fine Tuning}
%Add stuff.\footnote{In fact, in a non-supersymmetric theory, the situation would
%be more stable if these scalar $\OO_i$ were replaced with pseudoscalar
%ones, for which it is more natural that $\vev{\OO_i}=0$; one would
%then replace $F^2$ with $F\tilde F$, etc.  Our results would not
%change however.} 

\subsection{A comment on the far infrared}
\label{subsec:IR}

In general, conformal invariance in the hidden sector may not hold
down to arbitrarily low energy.  Indeed, we have just discussed
various ways in which conformal invariance may be violated at low
scales.  Moreover, with the couplings that we consider, a truly
conformal sector with very light particles can potentially induce new
processes that have not been observed, or affect big-bang
nucleosynthesis or other aspects of cosmology or astrophysics.  For
these reasons it may be that the hidden sector has a mass gap at some
scale $\mu$, which truncates all the branch cuts in Green functions of
hidden-sector operators.  (Examples of how this could occur appear in
\cite{unFox, unDelgado, hvun}.)  We will assume that any such $\mu$ is low
enough that (1) it does not impact hidden-sector Green functions above
a few tens of GeV, and (2) it does not cause any ``hidden valley''
signatures, where production of conformal excitations at high energy
turns into hidden particles at the scale $\mu$, which in turn decay to
standard model particles on detector time scales, giving visible
signatures \cite{hvun} and completely changing the LHC phenomenology.
{\it We assume throughout this paper that any infrared effects do not
affect the basic unparticle paradigm: that the hidden sector dynamics,
for all observable purposes at the Tevatron and LHC, is conformally
invariant and therefore predominantly invisible.}

\section{The bound applied to four-photon events}
\label{sec:boundon4gam}

 We now assume that the
Lagrangian has couplings between the two sectors of the form
\begin{equation}\label{somecouplings} {\cal L} = 
{1\over \Lambda_1^{\Delta_1}}\OO_1
\sum_a G^a_\mn G^{a\mn} + {1\over\Lambda_2^{\Delta_2}}\OO_2 F_\mn
F^{\mn} 
\end{equation}
where $G^a$ ($a=1,\dots,8$) and $F$ are $SU(3)$ and $U(1)$-electromagnetic
%hypercharge
field-strength tensors.  For consistency, since the events we will
study have energies far above the 100 GeV scale, we actually must
couple the operator $\OO_2$ to hypercharge bosons, with a coefficient
$(\Lambda_2^{\Delta_2}\cos^{2}\theta_W)^{-1}$.  But for brevity we
will ignore the associated $\gamma Z$ and $ZZ$ couplings for this
paper.  Although they contribute comparable three-photon and/or large
MET signals, including them would not change the bounds that we
obtain, which are in fact bounds on {\it the sum} of the
cross-sections for all these processes.  Thus this omission is
conservative, and simplifies our presentation.

Note that we make explicit that $\OO_1$ and
$\OO_2$ are distinct operators, potentially with $\Delta_1\neq
\Delta_2$ and $\Lambda_1\neq\Lambda_2$.  This need not be the case.
They might be distinct operators with $\Delta_1=\Delta_2$, or with
equal $\Lambda_i$.  Or we might take $\OO_1=\OO_2$, as was assumed in
\cite{FengRajTu}; in this case we could assume $\Lambda_1=\Lambda_2$,
as in \cite{FengRajTu}, but we need not do so.  
In this sense our analysis is more general than that
of \cite{FengRajTu}.  Indeed we will see the case they considered is
much more strongly constrained than is the general situation.

Now let us carry out our argument.  Suppose, as we will obtain in the
next section, that we have a lower bound on the scale $\Lambda_1$
for given $\Delta_1$.  This is then an upper bound on the cross-section
$\sigma(g g \to \{X\};\shat)$ for producing anything in the hidden
sector via the operator $\OO_1$.  
We could obtain from this
a bound on the total hadronic cross-section $\sigma(pp\to \{X\})$ by
convolving this bound with the gluon distribution function in the
proton.  But this is not our goal.

Instead, we turn to any particular process such as $gg\to
\gamma\gamma\gamma\gamma$, and require that it not be so large 
as to make preservation of conformal invariance impossible. In short,
we require 
\begin{equation}\label{bound4gam}
\sigma(gg\to\fourgamma;\shat) < \sigma(gg\to\{X\};\shat)
   \ \ \ \ (\shat < \shat_{max})
\end{equation}
But what $\sqrt{\shat_{max}}$ should we choose?

To choose $\sqrt{\shat_{max}}$ to be the collider energy would be too
strong a condition.  Most $gg\to \fourgamma$ events at any collider will
occur at energies far below the total collider energy, and so
$\sqrt{\shat_{max}}$ need not be nearly so high.  To determine the
appropriate energy, we must compute the four-photon cross-section as a
function of $\shat$, under the assumption of conformal invariance, and
see where it is large.  Then we should
choose $\shat_{max}$ so that the great majority of the $\fourgamma$
events will be produced at energies below this value.

For example, we might reasonably demand that a certain
fraction $\zeta$
%$\frac23$ 
 of the $gg\to
\fourgamma$ cross-section must occur below the scale
$\sqrt{\shat_{max}}$.  That is, we define $\shat_{max}$ by
\begin{equation}\label{definesmax1}
\int_0^{\shat_{max}} d\shat\ \frac{d\sigma(gg\to \fourgamma)}{d\shat} = 
%\frac23
\zeta \int_0^{s} d\shat\ \frac{d\sigma(gg\to \fourgamma)}{d\shat}
\end{equation}
where $s$ is the square of the collider center-of-mass energy. 
To require $\zeta=1$, and therefore $\shat_{max}=s$, 
would be far too strong, as noted above.
If we instead took $\zeta=\frac12$
% of the $gg\to
%\fourgamma$ cross-section occur below the scale $\sqrt{\shat_{max}}$,
then we would effectively be demanding, typically, that the peak
cross-section for $gg\to\fourgamma$ occurs at $\shat_{max}$, right
where conformal invariance is breaking down.  In this case, none of
the predictions (cross-section or kinematic distributions) of
\cite{FengRajTu} would be at all reliable.  For this reason we view
$\zeta=\frac12$ as unreasonable.  We therefore take $\zeta=\frac23$ as
a conservative choice.  This should ensure that the prediction for the
rate and differential distributions for $gg\to\fourgamma$ are given to
a rough approximation by conformally invariant calculations, and are
not beset with model-dependent effects beyond roughly the 30\%--50\%
level.

Importantly, assuming {\it only} that conformal invariance has not
been violated, we can determine $\shat_{max}$ in a completely
model-independent way that depends only on $\Delta_1$ and $\Delta_2$.
From \Eref{ABKKKKsigma} (with $\Delta_1=\delta_1$ and
$\Delta_3=\Delta_2=\delta_2$ in the case at hand), we know the precise
$\shat$ dependence of the cross-section, up to constants that
factor out of the condition in \Eref{definesmax1}.  Defining the $gg$
luminosity function as usual by 
\begin{equation}\label{gglum}
\frac{dL_{gg}(\tau)}{d\tau} = \int dy f_g(\sqrt\tau e^y) f_g(\sqrt\tau e^{-y})
\end{equation}
(where $\tau=\hat s/s$)
and substituting from \Eref{ABKKKKsigma}, we have, for $\zeta=\frac23$,
%\begin{equation}\label{definesmax2a}
%\frac{1}{s}\int_0^{\shat_{max}} d\shat\ 
%\frac{dL_{gg}(\shat/s)}{d(\shat/s)}
%{\shat^{\Delta_1+2 \Delta_2-1}}=  
%\frac{2}{3}\
%\frac{1}{s}\int_0^{s} {d\shat}\ 
%\frac{dL_{gg}(\shat/s)}{d(\shat/s)}
%{\shat^{\Delta_1+2 \Delta_2-1}} \ ,
%\end{equation}
%%
%or more simply,
\begin{equation}\label{definesmax2}
\int_0^{\shat_{max}/s} d\tau
\frac{dL_{gg}(\tau)}{d\tau}
{\tau^{\Delta_1+2 \Delta_2-1}}=  
\frac{2}{3}
\int_0^{1} {d\tau}\ 
\frac{dL_{gg}(\tau)}{d\tau}
{\tau^{\Delta_1+2 \Delta_2-1}}
\end{equation}
Notice that all dependence on $C_{122}$,
$N_{\fourgamma}(\Delta_1,\Delta_2)$ and $\Lambda_i$ factors out of
this expression.  Thus our choice of $\shat_{max}$, once we have
chosen a fixed $\zeta$, depends only on $\Delta_1+2\Delta_2$, and
largely scales with $s$ (up to the slow variation of $L_{gg}$ through the
evolution of the gluon distribution function.)  
Table \ref{table:smax} shows $\sqrt{\shat_{max}}$
for a 10 TeV LHC and various choices of $\Delta_1+2\Delta_2$.

At this point we should mention that throughout this paper our
numbers are produced using the (outdated)
CTEQ5M pdfs \cite{CTEQ5}.  This is purely for technical reasons of
calculational speed.  Results obtained from more modern pdfs differ by
significantly less than other systematic errors in our calculations.
We have explicitly checked in several cases that our numbers do not
change significantly with the
MSTW08 pdf set \cite{MSTW8}.  The errors on $\shat_{max}$ from
uncertainties in the gluon pdfs and the appropriate choice of
factorization scale are estimated at approximately 5 percent.  This is
smaller than the dominant source of uncertainty, which arises
from the choice of $\zeta$ that defines $s_{max}$.  We will have more
to say about this uncertainty after we present our results.
% Table created by WinTeX XP: 2 Columns x 7 Rows.
\begin{table}\centering
\begin{tabular}{|l||c|c|c|c|c|c|c|}
\hline
$\Delta_1+2\Delta_2$&          3.0 & 3.5 & 4.0 & 4.5 & 5.0 & 5.5 & 6.0 \\
\hline
$\sqrt{\shat_{max}}$ (in TeV)  &1.2 &1.7 & 2.2 & 2.7 & 3.1 & 3.4 & 3.7 \\ 
\hline 
%Row: 1
%\cline{1-2}
%\vbox to1.88ex{\vspace{1pt}\vfil\hbox to22.80ex{\hfil $\Delta_1+2\Delta_2$\hfil}} & 
%\vbox to1.88ex{\vspace{1pt}\vfil\hbox to25.20ex{\hfil $\sqrt{\shat_{max}}$ (in TeV)\hfil}} \\
%\cline{1-2}
%\vbox to0.28ex{\vspace{.0pt}\hbox to2.80ex{\hfil \hfil}} & 
%\vbox to0.28ex{\vspace{.0pt}\hbox to2.20ex{\hfil \hfil}} \\
%Row: 2
%\cline{1-2}
%\vbox to1.88ex{\vspace{1pt}\vfil\hbox to22.80ex{\hfil 3\hfil}} & 
%\vbox to1.88ex{\vspace{1pt}\vfil\hbox to25.20ex{\hfil 1.2\hfil}} \\
%Row: 3
%\cline{1-2}
%\vbox to1.88ex{\vspace{1pt}\vfil\hbox to22.80ex{\hfil 3.5\hfil}} & 
%\vbox to1.88ex{\vspace{1pt}\vfil\hbox to25.20ex{\hfil 1.7\hfil}} \\
%Row: 4
%\cline{1-2}
%\vbox to1.88ex{\vspace{1pt}\vfil\hbox to22.80ex{\hfil 4\hfil}} & 
%\vbox to1.88ex{\vspace{1pt}\vfil\hbox to25.20ex{\hfil 2.2\hfil}} \\
%Row: 5
%\cline{1-2}
%\vbox to1.88ex{\vspace{1pt}\vfil\hbox to22.80ex{\hfil 4.5\hfil}} & 
%\vbox to1.88ex{\vspace{1pt}\vfil\hbox to25.20ex{\hfil 2.7\hfil}} \\
%%Row: 6
%\cline{1-2}
%\vbox to1.88ex{\vspace{1pt}\vfil\hbox to22.80ex{\hfil 5\hfil}} & 
%\vbox to1.88ex{\vspace{1pt}\vfil\hbox to25.20ex{\hfil 3.1\hfil}} \\
%%Row: 7
%\cline{1-2}
%\vbox to1.88ex{\vspace{1pt}\vfil\hbox to22.80ex{\hfil 5.5\hfil}} & 
%\vbox to1.88ex{\vspace{1pt}\vfil\hbox to25.20ex{\hfil 3.4\hfil}} \\
%Row: 8
%\cline{1-2}
%\vbox to1.88ex{\vspace{1pt}\vfil\hbox to22.80ex{\hfil 6.0\hfil}} & 
%\vbox to1.88ex{\vspace{1pt}\vfil\hbox to25.20ex{\hfil 4.1\hfil}} \\
%\cline{1-2}
\end{tabular}
\caption{Values of $\sqrt{\shat_{max}}$, at a 10 TeV LHC, 
for various choices of $\Delta_1+2\Delta_2$. }
\label{table:smax}
\end{table}

Now let us return to the process of obtaining a bound.
The bound arises from the fact that
$\sigma(gg\to\{X\};\shat)$ is precisely
known, except for an overall constant normalization, which depends
only on $\Lambda_1$ and is proportional to $1/\Lambda_1^{2\Delta_1}$.
If $\Lambda_1$ is bounded from below, $\Lambda_1 >
\Lambda_1^{min}$, then $\sigma(gg\to\{X\};\shat)$ is likewise bounded
from above, at all $\shat$, by $\sigma(gg\to\{X\};\shat;
\Lambda_1^{min})$.

\begin{figure}
\begin{minipage}[b]{0.45\textwidth} % A minipage that covers half the page
\centering
\includegraphics[width=4.5cm]{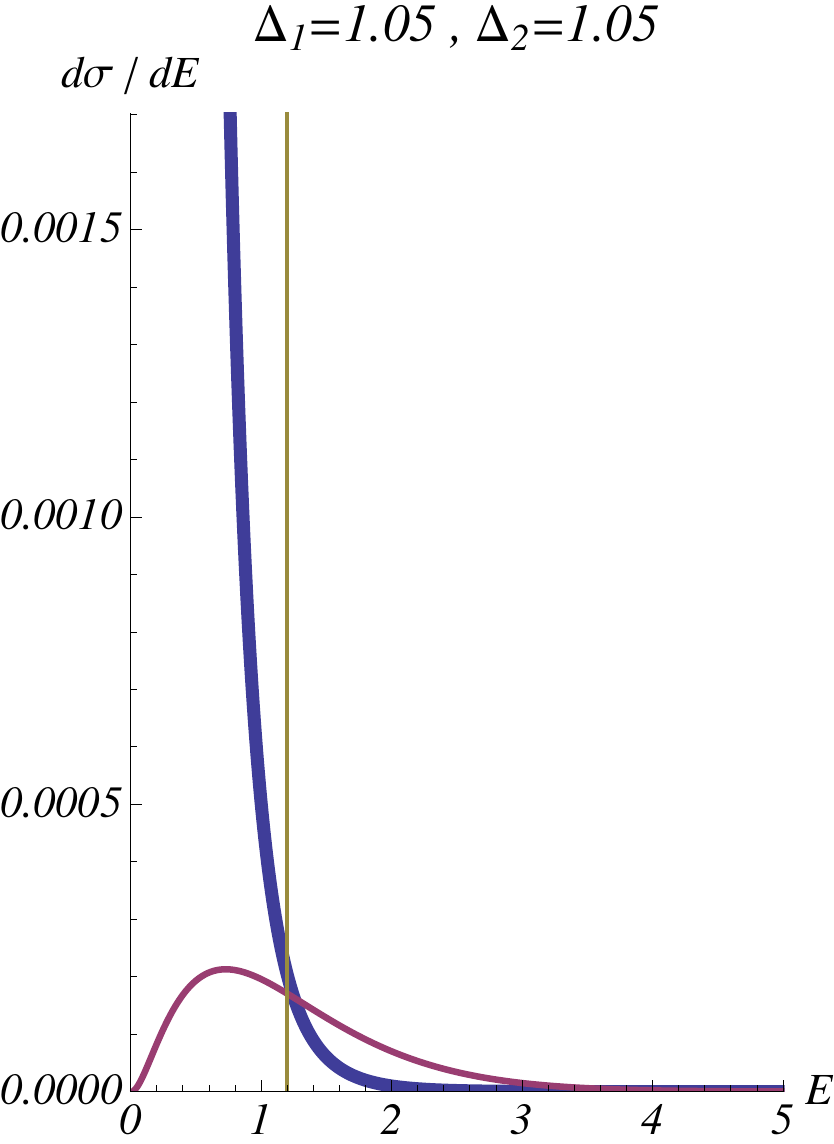}
\end{minipage}
\hspace{0.5cm} % To get a little bit of space between the figures
\begin{minipage}[b]{0.45\textwidth}
\centering
\includegraphics[width=4.5cm]{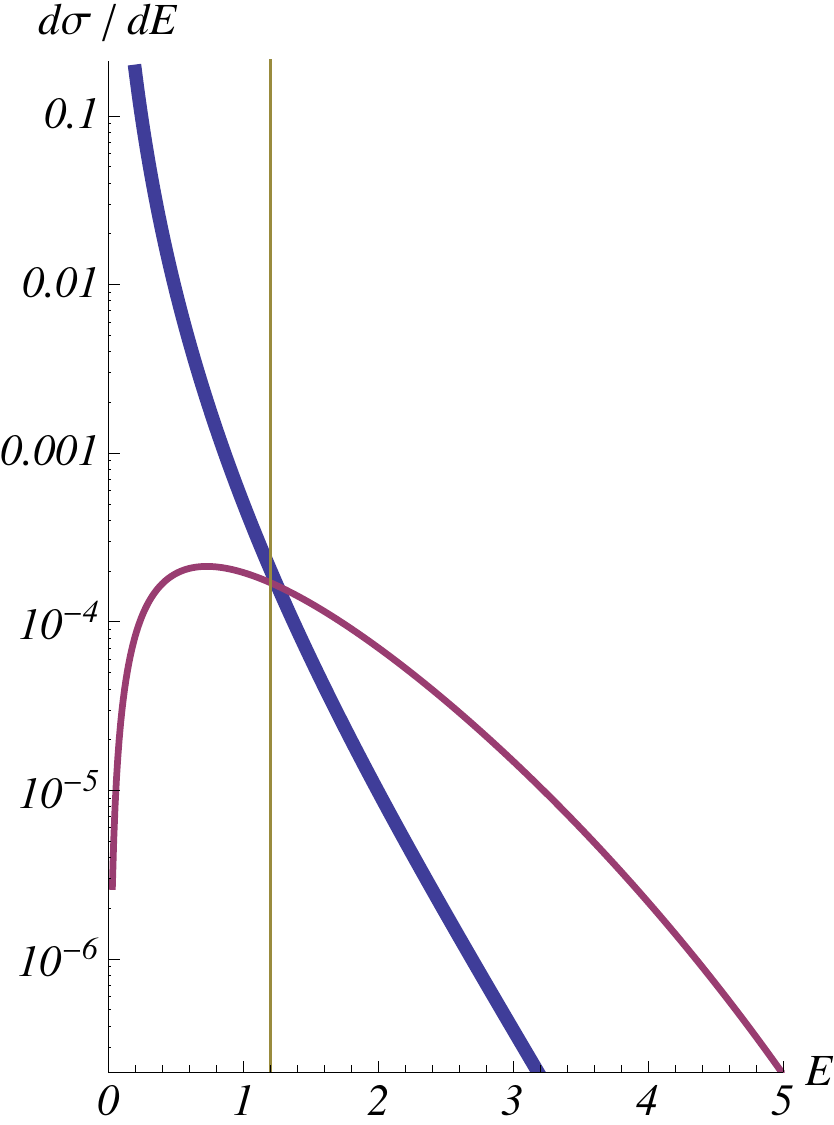}
\end{minipage}
\caption{For $\Delta_1=1.05,\Delta_2=1.05$, the differential
 cross-sections (in pb/GeV) versus energy $E=\sqrt{\shat}$
(in TeV) for all production
 processes involving the hidden sector (thick curve) and for
 four-photon production (thin curve). The right hand plot is the
 same as the left hand plot, but on a log scale. The total hidden-sector
 cross-section is normalized by our bound on $\Lambda_1$, and the
 four-photon cross-section is normalized so that it satisfies
 unitarity, by not exceeding the total for any $\shat<\shat_{max}$.
 Our estimate of $\sqrt{\shat_{max}}$, determined as explained in the
 text, is indicated by the vertical line.   }
\label{fig:fig00}
\end{figure}

\begin{figure}
\begin{minipage}[b]{0.45\textwidth} % A minipage that covers half the page
\centering
\includegraphics[width=4.5cm]{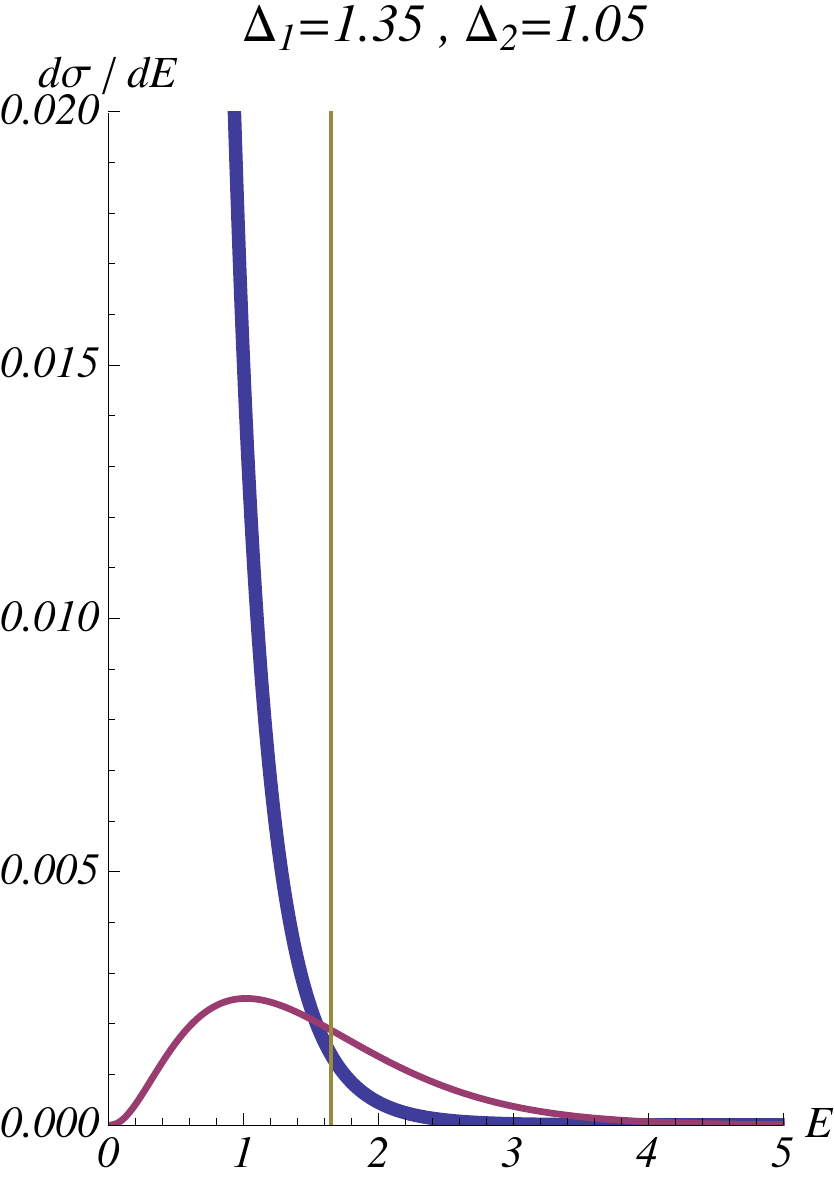}
\end{minipage}
\hspace{0.5cm} % To get a little bit of space between the figures
\begin{minipage}[b]{0.45\textwidth}
\centering
\includegraphics[width=4.5cm]{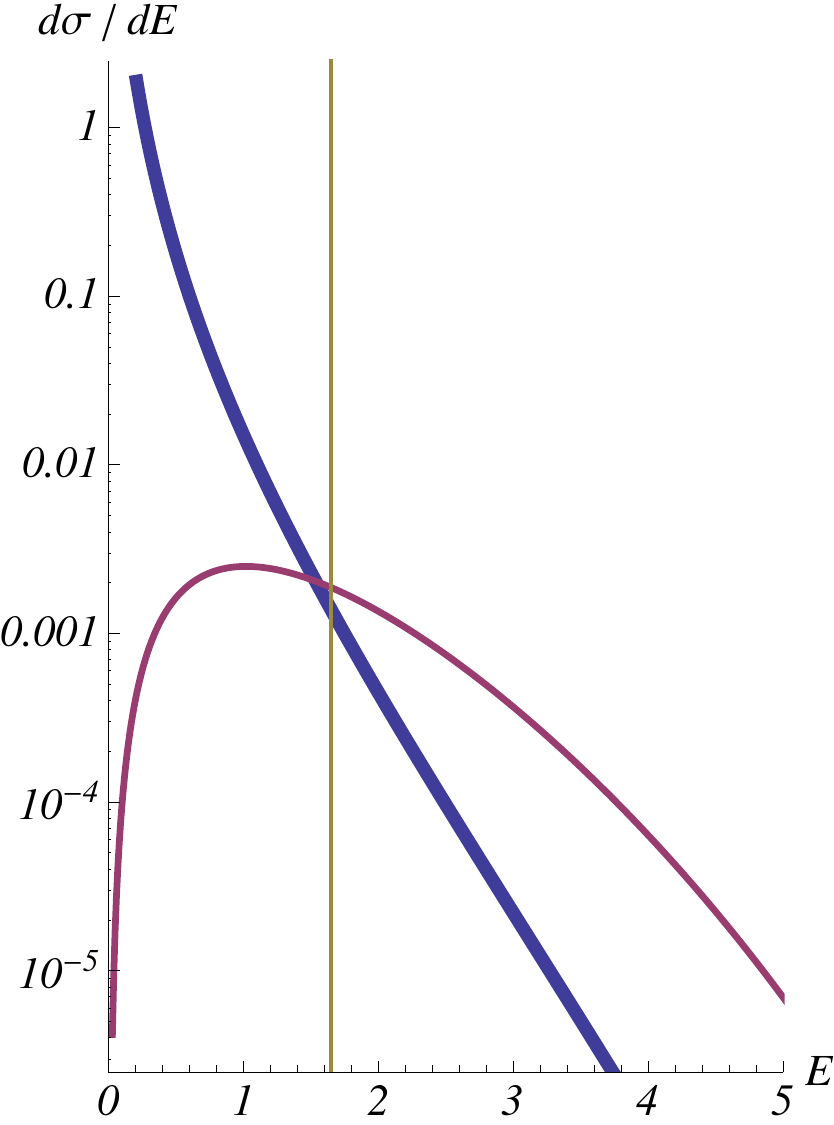}
\end{minipage}
\caption{Same as \reffig{fig00}, but with $\Delta_1=1.35,\Delta_2=1.05$.   }
\label{fig:fig30}
\end{figure}

\begin{figure}
\begin{minipage}[b]{0.45\textwidth} % A minipage that covers half the page
\centering
\includegraphics[width=4.5cm]{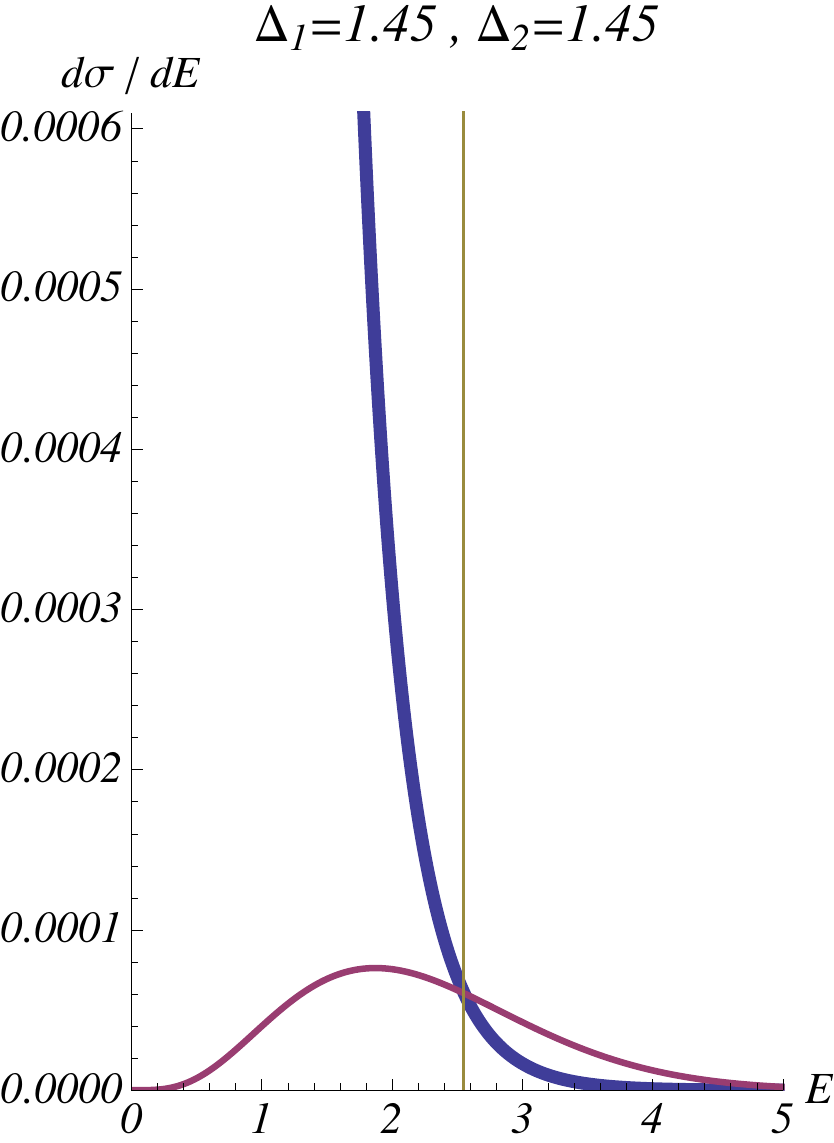}
\end{minipage}
\hspace{0.5cm} % To get a little bit of space between the figures
\begin{minipage}[b]{0.45\textwidth}
\centering
\includegraphics[width=4.5cm]{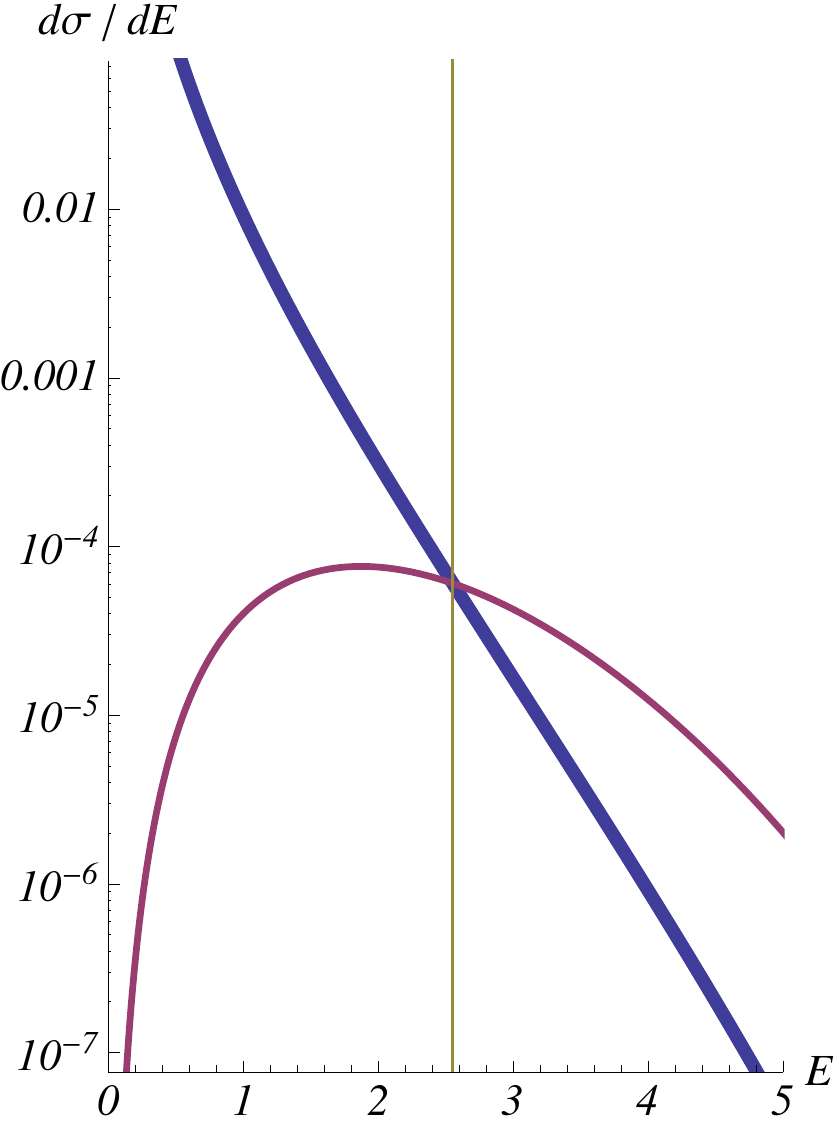}
\end{minipage}
\caption{Same as \reffig{fig00}, but with $\Delta_1=1.45,\Delta_2=1.45$.   }
\label{fig:fig44}
\end{figure}

\begin{figure}
\begin{minipage}[b]{0.45\textwidth} % A minipage that covers half the page
\centering
\includegraphics[width=4.5cm]{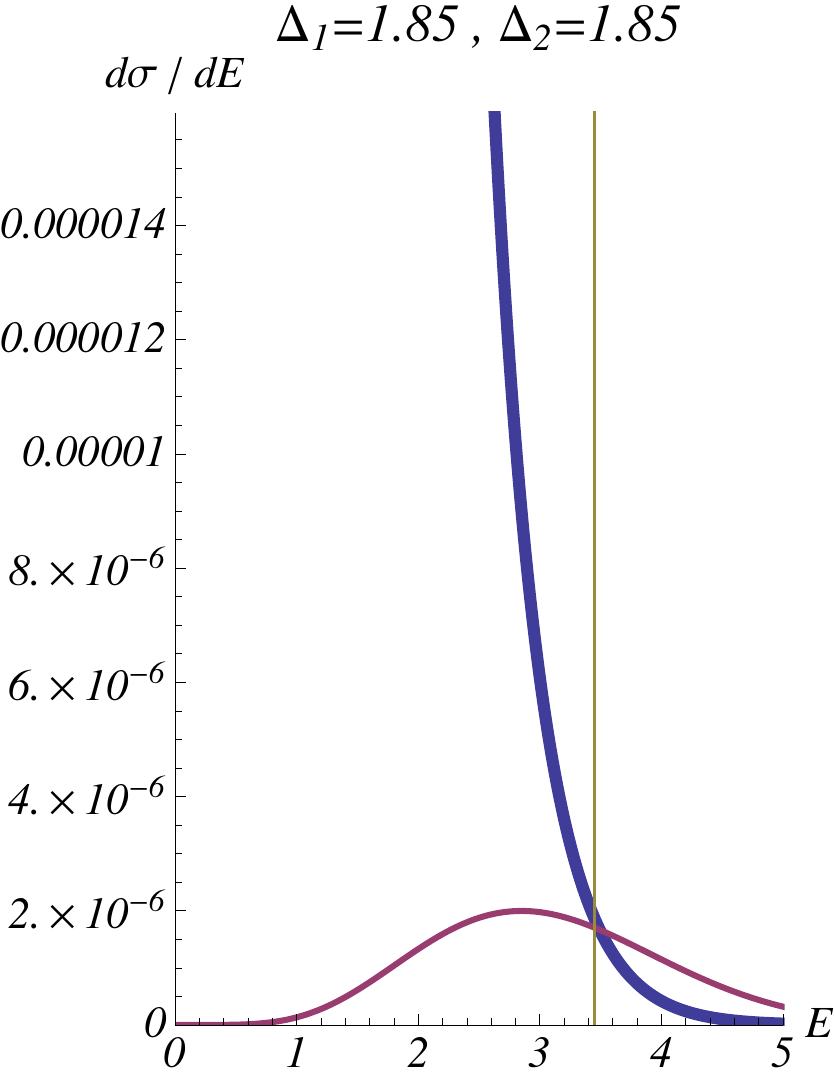}
\end{minipage}
\hspace{0.5cm} % To get a little bit of space between the figures
\begin{minipage}[b]{0.45\textwidth}
\centering
\includegraphics[width=4.5cm]{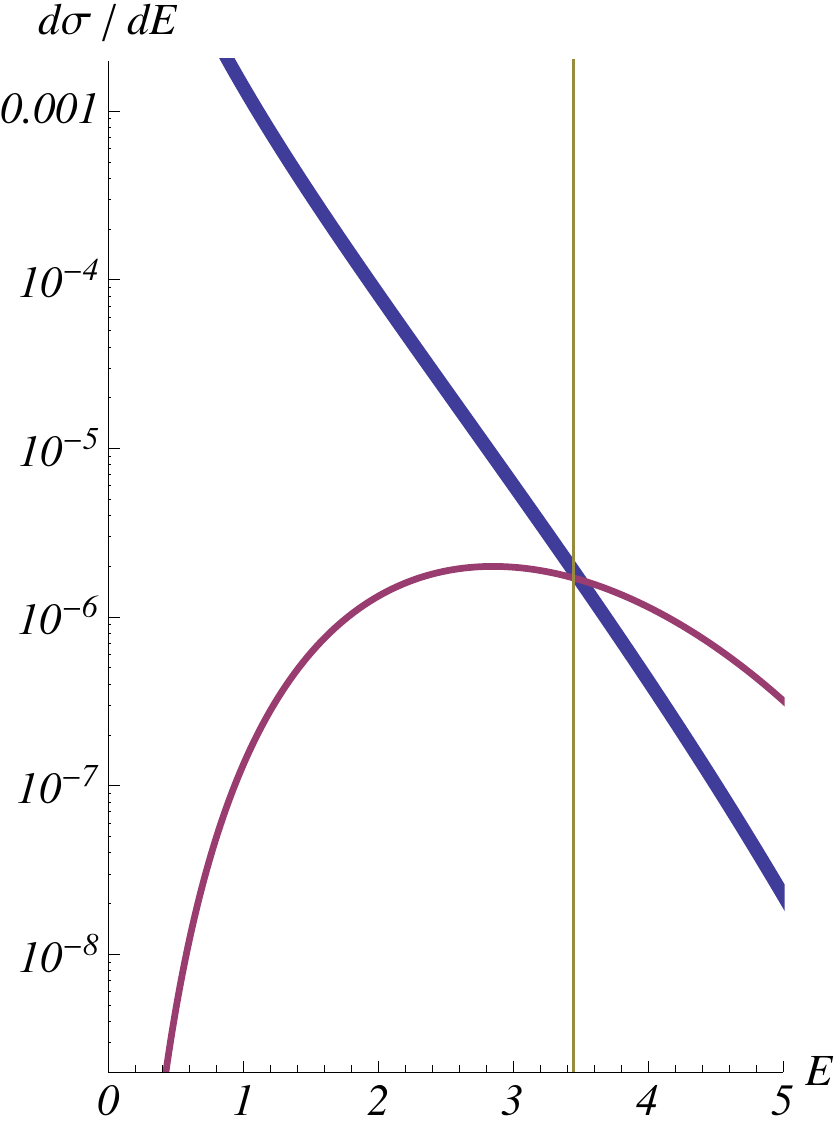}
\end{minipage}
\caption{Same as \reffig{fig00}, but with $\Delta_1=1.85,\Delta_2=1.85$.   }
\label{fig:fig882}
\end{figure}

To understand what this means intuitively, we have plotted
$\sigma(gg\to \fourgamma;\shat;\Lambda_1^{min})$ and
$\sigma(gg\to\{X\};\shat)$ in figures \ref{fig:fig00} to
\ref{fig:fig882}, for several different choices of $\Delta_1$ and
$\Delta_2$.  The total hidden-sector cross-section $\sigma(gg\to
\fourgamma;\shat;\Lambda_1^{min})$ is normalized to saturate the bound
on $\Lambda_1$ that we will obtain later; however for the moment the
shape matters more than the normalization.  The normalization of the
$\fourgamma$ cross-section $\sigma(gg\to\{X\};\shat)$ is chosen so
that it does not exceed the total hidden-sector cross-section at any
$\sqrt{\shat}$ below $\sqrt{\shat_{max}}$, whose value is indicated by
a vertical line.  Because of the rate with which the $gg$ luminosity
decreases, $\sigma(gg\to \fourgamma;\shat)$ initially increases with
energy, until the rapid decrease of the $gg$ luminosity at high
$\shat$ overwhelms the rising partonic cross-section.  Meanwhile,
$\sigma(gg\to\{X\};\shat)$ decreases rapidly everywhere.  Because of
this, we can see by eye that $\shat_{max}$ must be taken quite large,
typically of order 1--4 TeV.  (This confirms that for $gg\to
\fourgamma$ we can neglect any effects from an infrared scale $\mu$ of
the sort discussed in Sec.~\ref{subsec:IR}.)  Also, we can see by eye
that $\sigma(gg\to \fourgamma;\shat)$ is always vastly less than
$\sigma(gg\to\{X\};\shat)$, because of the shapes of the two curves,
until $\shat$ is very close to $\shat_{max}$.  

As we noted earlier in our more general discussion, dimensional analysis
always assures that the ratio
$\sigma(gg\to \fourgamma;\shat)/\sigma(gg\to\{X\};\shat)$ 
grows with energy, as long as 
conformal invariance is applicable.  Therefore
\begin{equation}\label{inequalities}
\frac{\sigma(gg\to \fourgamma;\shat)}{\sigma(gg\to\{X\};\shat)}
<
\frac{\sigma(gg\to \fourgamma;\shat_{max})}{\sigma(gg\to\{X\};\shat_{max})}
<
\frac{\sigma(gg\to \fourgamma;\shat_{max})}
{\sigma(gg\to\{X\};\shat_{max};\Lambda_1^{min})} \ .
\end{equation}
Unitarity requires the last expression be less than one,
%\begin{equation}\label{simplebound1}
%\sigma(gg\to \fourgamma;\shat_{max}) \ll
%{\sigma(gg\to\{X\};\shat_{max};\Lambda_1^{min})} \ 
%\end{equation}
and writing this condition in terms of the constant
coefficients appearing in the formulas \eref{sigmatot} and \eref{ABKKKKsigma} 
for the cross-sections, we obtain
\begin{equation}\label{boundoncoeffs}
|C_{123}|^2N_{\fourgamma}(\Delta_1,\Delta_2)\Lambda_2^{-4\Delta_2}\ll
\left({\shat_{max}}\right)^{-2\Delta_2}N_0(\Delta_1) 
\ .
\end{equation}
Notice all $\Lambda_1$ dependence factors out of this bound.

Finally we may obtain a bound on the total cross-section
for $gg\to\fourgamma$, namely
\beq\label{finalbound}
\sigma(pp\to \fourgamma)
&=&|C_{123}|^2
N_{\fourgamma}(\Delta_1,\Delta_2)
\Lambda_1^{-2\Delta_1}\Lambda_2^{-4\Delta_2}
{s^{\Delta_1+2 \Delta_2-1}} 
\int_0^{1} d\tau \frac{dL_{gg}(\tau)}{d\tau}
{\tau^{\Delta_1+2 \Delta_2-1}} \nonumber \\
%\frac{1}{s}\int_0^{s} d\shat\ \frac{dL_{gg}(\shat/s)}{d(\shat/s)}
%{\shat^{\Delta_1+2 \Delta_2-1}} \nonumber \\
&\ll&
N_0(\Delta_1)\
\Lambda_1^{-2\Delta_1} 
\left({\shat_{max}}\right)^{-2\Delta_2}
{s^{\Delta_1+2 \Delta_2-1}} 
\int_0^{1} d\tau\ \frac{dL_{gg}(\tau)}{d\tau}
{\tau^{\Delta_1+2 \Delta_2-1}}\nonumber \\
%\frac{1}{s}\int_0^{s} d\shat\ \frac{dL_{gg}(\shat/s)}{d(\shat/s)}
%{\shat^{\Delta_1+2 \Delta_2-1}}\nonumber \\
&<&
\frac{N_0(\Delta_1)}{s}\
\left(\frac{{s}}{[\Lambda_1^{min}]^2}\right)^{\Delta_1}
\left(\frac{s}{\shat_{max}}\right)^{2\Delta_2}
% {s^{\Delta_1+2 \Delta_2-1}} 
\int_0^{1} d\tau\ \frac{dL_{gg}(\tau)}{d\tau}
{\tau^{\Delta_1+2 \Delta_2-1}}
%\frac{1}{s}\int_0^{s} d\shat\ \frac{dL_{gg}(\shat/s)}{d(\shat/s)}
%{\shat^{\Delta_1+2 \Delta_2-1}}%
%\cr
%&\ &\
\eeq
This is the formal expression of our main result.

Notice that our bound only depends on the collider energy $s$,
on the dimensions $\Delta_1$ and $\Delta_2$,
on $\shat_{max}/s$ (determined using \Eref{definesmax2} by
$\Delta_1$ and $\Delta_2$,) on $N_0$ (which 
is 
\begin{equation}
N_0(\Delta_1) = %|\Imm \big[G_\OO^{(0)}(q)\big] =% |f_{gg}|^2
\frac{-\sin(\pi\Delta_1)}{(4 \pi)^{2\Delta_1-2}} 
\frac{\Gamma[2-\Delta_1]}{\Gamma[\Delta_1]} \ 
\end{equation}
for a $gg$ initial state,)
on the known $gg$ luminosity, and finally on
$\Lambda_1^{min}$ (which we must separately determine using
theoretical and experimental constraints.)  {\it All dependence on
$\Lambda_2$, $N_{\fourgamma}$ and $C_{122}$ has vanished.}  If we know
$\Lambda_1^{min}$ as a function of $\Delta_1$ and perhaps $\Delta_2$,
we can obtain a bound that is model-independent and depends only on
$\Delta_1$ and $\Delta_2$.

\section{Obtaining bounds on $\Lambda_1$}

Our only remaining task is to determine $\Lambda_1^{min}$.  Once we
have it, we can compute the bound on the $gg\to \fourgamma$
cross-section.

We apply two main considerations for constraining $\Lambda_1$.  The
first is that if $\Lambda_1$ is too low, then not only is the rate for
the invisible process $\sigma(pp\to\{X\})$ very large, the observable
process $\sigma(pp\to j +\{X\})$, where $j$ is an initial-state jet,
becomes comparable to the standard model rate for jet plus
missing transverse momentum (MET).  Contraints from Tevatron, mainly
from the CDF study \cite{CDFjMET}, put strong contraints on
$\Lambda_1$ for low $\Delta_1$.

A second constraint on $\Lambda_1$ comes from the fact that the
coupling of $\OO$ to gluons {\it itself} induces corrections to
$G_\OO^{(0)}$.  We must assume these are small if we are to use
conformal invariance to make predictions regarding $gg\to\fourgamma$.
Either such predictions are impossible, invalidating the approach of
\cite{FengRajTu}, or $\Lambda_1$ must be larger than some minimum.
This puts moderate constraints, of order $1.5$ TeV or larger,
which are relevant for larger
$\Delta_1$ where the experimental constraints are weakest.

\subsection{Bounds from Tevatron measurements of monojet events}

Given a known partonic cross-section for a hidden-sector process, it
is straightforward to compute the rate for jets-plus-MET 
where the jet(s) only arise from the initial state.  One might ask
whether emission from the final state could possibly compete with, and
perhaps interfere with, this process.  The answer regarding
interference is ``no''; once the hidden state has been produced, it is
color neutral, and any final-state radiation must be color singlet,
requiring at least two jets be emitted.  Similarly, in the model we are
considering, the largest interactions between the two sectors involve
irrelevant couplings, so any final state radiation process is small
at low energy, and is either too small to observe or would show up as a
large tail at high energy.  Since no such tail is observed at Fermilab, we 
assume any final-state radiation of jets cannot affect the limits which
we will now obtain.

For a conformal hidden sector produced through $gg$, the rate is
entirely fixed by $\Lambda_1$ and $\Delta_1$. For $qg\to q\{X\}$, we
find, at leading order, 
\begin{equation}\label{jetMETrate}
 \frac{d\sigma}{dp_T^2}(qg\to q + \{X\}) = 
C {\sum_{n=0}^3} B_n\times
{{ }_2F_1}\left[\frac {1} {2}, -1 + \Delta_1+n, -\frac {1} {2} + \Delta_1+n,
\frac {\shat - 2\sqrt {\shat} {p_T}} {\shat + 2\sqrt {\shat}
{p_T}} \right] \ ,
\end{equation}
where $p_T$ is the transverse momentum of the jet,
\be C= \frac{4\pi^2 {\alpha_s} }{(2{\pi}
{\Lambda} )^{2 \Delta_1} }
\frac{ \Delta_1}{3 (3 + 2 \Delta_1) {\Gamma}[2 + 2 \Delta_1]} 
\frac{ ({\shat} - 2{p_T} \sqrt{\shat})^{-1 + \Delta_1}}
{{p_T}^2 {}\sqrt {(1 - 4 {p_T}^2/ {\shat})}}
\end{equation}
and
\beq
  B_0 &=& \left
(-3 - 2 \Delta_1 + 12 \Delta_1^2 + 8 \Delta_1^3 \right) (2 - 3 {p_T}^2/ {\shat})
\\
 B_1 &=&  - 2
(\Delta_1 - 1)\left (3 + 8 \Delta_1 + 4 \Delta_1^2 \right) 
(4 - 3 {p_T}^2/ {\shat})\left (1 - \sqrt {4 {p_T}^2/ {\shat}}
\right)
\\
  B_2 &=&  12 (\Delta_1
- 1) \Delta_1 {} (3 + 2 \Delta_1)\left (1 - \sqrt {4 {p_T}^2/ {\shat}} \right)^2
\\
  B_3 &=&  - 8
(\Delta_1 - 1) \Delta_1 {} (\Delta_1 + 1)
\left (1 - \sqrt {4 {p_T}^2/ {\shat}} \right)^3
%$$  {HG}[ {\Delta},  {s},  {p_T}] = 
%{Hypergeometric2F1}\left[\frac {1} {2}, -1 + \Delta, -\frac {1} {2} + \Delta,
%\frac {s - 2\sqrt {s {p_T}^2}} {s + 2\sqrt {s
%{p_T}^2}} \right]
%$$
\eeq 
(The reader may compare our result with the literature, see for
example \cite{Dawson}, in the $\Delta_1\to 1$ limit.)  This is the
dominant process at high energy at the Tevatron.  There is also the
process $gg\to g\{X\}$, but this is smaller in the energy range of
interest at the Tevatron and we neglect it.  If we included it, our
lower bounds on $\Lambda_1$ would be stronger.

The CDF experiment \cite{CDFjMET} has published results on monojet
events, in the context of a search for extra dimensions, and a public
webpage with additional information and plots is available
\cite{CDFjMETweb}.  Early results from DZero \cite{D0jMET}, with much
lower statistics, have not been updated; we will not use them in our
analysis.  The CDF study uses two sets of cuts, a loose set for a
model-independent search, and a tighter set optimized for an
extra-dimensions search; we use the former.  The data is available in
plot form, though not in table form; we have extracted the data
directly from the plots, introducing a moderate amount of systematic
error in the process.  Demanding that the process $qg\to q\{X\}$ not
be easily visible above the error bars of the plots in \cite{CDFjMET}
puts a limit on $\Lambda_1$ for any given $\Delta_1$.

Through this requirement we find limits on $\Lambda_1$ shown in
boldface in Table \ref{table:boundLambda1}.  There are substantial
systematic error bars on our results.  First, we have not included the
K-factor from loop corrections, or the process $gg\to g\{X\}$; doing so
would give a slightly stronger bound.  Second, we are not able to
include experimental efficiencies and effects of jet energy scale
uncertainties; doing so would give a slightly weaker bound.
Furthermore, our computation is done at leading order, for which the
jet transverse momentum $p_t^{jet}$ and the MET are equal.  However,
both additional jet radiation and jet mismeasurements contribute in
the data, so these are not in fact equal, and thus when we extract a
limit on $\Lambda_1$ it is inherently ambiguous whether we should use
the experimental distributions of $d\sigma/d({\rm MET})$ or
$d\sigma/d(p_t^{jet})$ (and neither is accurate beyond leading order.)
%This being said, our limits are extracted from a relatively
%high $p_T$ region ($p_t^{jet}>150$ GeV) so this ambiguity is not
%particularly large.  
Crudely, we estimate that the errors on our
determination of $\Lambda_1$ are of order 10 percent, which turns out
to be a subleading uncertainty compared to that
stemming from the ambiguity in choosing $\shat_{max}$.

As final comments, we note that for $p_t^{jet}$ of this size, the
cross section for $d\sigma/dp_t^{jet}$ involves an integral over $q^2$
that is insensitive to low $q^2$.  In other words, our limits on
$\Lambda_1$ are insensitive to any low-energy cutoff $\mu$.  Also, the
reader may observe that our calculations do not suffer from the
well-known singularity at $\Delta\to 2$ which indicates the need for
renormalization.  This is because our results depend only on the
imaginary part of $G_\OO$.  All of our results are smooth as $\Delta$
passes through 2.

%\begin{table}[b]
%\begin{tabular}[c]{|| c | c |c||}\hline
%\ $\Delta_1$ \ &\   $\Lambda_1^{min}$ (in TeV) from $j$+MET
% &\   
%$\Lambda_1^{min}$ (in TeV) from conformal invariance below $\shat_{max}$
%\\
%\hline
% \ & \ & $\Delta_2=$1 \ \ \ $|$ \ \ \ $\Delta2=2$ \\
%\ 1.1\ &\ 10 TeV \ & 1.4 \ \ \ | \ \ \ 3.2 TeV \\
%% 1       13.6
%\ 1.25 \ &    3.3 TeV \ & 1.8 -- 3.8 TeV \\
%\ 1.5  \ &    1.2 TeV \ & 1.7 -- 3.2 TeV  \\
%\ 1.75  \ &    0.7 TeV \ & 1.5 -- 2.5 TeV \\
%\ 2   \ &    0.4 TeV \ & 1.3 -- 2.1 TeV \\
%\hline 
%\end{tabular}
%\caption{The minimum values of $\Lambda_1$, as a function of $\Delta_1$, allowed
%by the experimental constraints from monojets and by the theoretical
%constraint that conformal invariance be preserved below $\shat_{max}$ for the 
%corresponding $\Delta_1$; see Table \ref{table:smax}.}
%\label{table:boundLambda1}
%\end{table}

\subsection{Bounds from preserving conformal invariance}

We noted earlier that in a conformal theory perturbed by an interaction
of the form \Eref{coupling1}, there is an irreducible effect that causes
$G_\OO(q;\Lambda)$ to differ from its conformal form
$G_\OO^{(0)}(q)$, given by \Eref{resummed2OO} and shown in \reffig{GSig}.  

At leading order, the QCD interactions of gluons play no role,
and so we may treat them as a system of free
massless particles --- a conformal field theory.  Thus our calculation
is a specific example of a more general issue: if we have two
conformal field theories $I$ and $J$, and we couple them through an
irrelevant operator $\OO_I \OO_J$ with coupling
$1/\Lambda^{4-\Delta_I-\Delta_J}$, where $\OO_I$ ($\OO_J$) is a scalar
operator in conformal sector $I$ ($J$), then this coupling leads formally
to a bad breaking of conformal invariance at some high scale
$M_{max}$.  More precisely, either conformal invariance is badly
broken, or the pointlike coupling $\OO_I \OO_J$ develops a
non-pointlike structure due to new physics at some scale at or below
$M_{max}$.  Either way, the approximation that one has two conformal
field theories coupled by a pointlike operator must break down.

What is an estimate for $M_{max}$?  With conventionally normalized
operators $\OO_I$ and $\OO_J$ one might naively guess through naive
dimensional analysis that $M_{max}\sim 4 \pi \Lambda$.  With the
normalization used in the unparticle literature (which sets the
conventions for our definition of $\Lambda_i$ in this paper), this is
essentially correct.  

However, the standard model operator $\sum_a G_\mn^a G^{a\mn}$ is {\it
not} a conventionally normalized operator of dimension 4, because it
contains derivatives.  One may easily check that these produce
additional factors of $2\pi$ (just as is expected in naive dimensional
analysis) leading to a $(2\pi)^4$ enhancement relative to the two
point function of a conventionally-normalized operator of dimension 4.
In addition, there is a factor of $8=3^2-1$ from the sum over colors.
Altogether this means that, for the normalization of $\Lambda_1$ given
through the use of the action \Eref{somecouplings}, which is the same
as used by Feng et al. in \cite{FengRajTu}, the breakdown of conformal
invariance occurs well below $4\pi\Lambda_1$.  This is significant
because in the literature one often sees discussion of taking
$\Lambda_1\sim 1$ TeV, which may cause conformal invariance to break
down within the range of energies accessible at LHC.  For our current
problem, since the peak of the $gg\to\fourgamma$ cross-section occurs
at energies typically greater than 1 TeV (see Table \ref{table:smax} and
Figs.~\ref{fig:fig00} -- \ref{fig:fig882}), this problem is
severe.

More precisely, the momentum-space two-point function of $G_\mn G^\mn$
is quartically divergent, and there are underlying quadratic and
logarithmic terms; renormalization removes these divergences but
leaves their finite contribution ambiguous.  However the imaginary
part of the two-point function is unambiguous, arising from a finite
$q^4 \ln q$ term.  When this imaginary part makes an order-one
correction to $G_\OO^{(0)}(q)$, conformal invariance is unambiguously
breaking down.

Even more precisely, we can see from \Eref{resummed2OO} that we can no
longer trust conformal invariance once $|G_\OO^{(0)}(q)\Sigma(q)|$ is of
order 1.  As we have just noted $\Sigma(q)$ is subject to
renormalization ambiguities, and for the same reason, so is
$G_\OO^{(0)}(q)$ if $\Delta_\OO\geq 2$.  But the imaginary parts of
$\Sigma$ and $G_\OO^{(0)}$ are not subject to such ambiguities.  
Noting
\begin{equation}
|G_\OO^{(0)}(q)\Sigma(q)| > 
\bigg|\Imm \big[G_\OO^{(0)}(q)\big] 
\Imm\big[\Sigma(q)\big]\bigg| 
\ ,
\end{equation}
we choose to apply an extremely conservative consistency condition,
namely
\begin{equation}
\bigg|\Imm \big[G_\OO^{(0)}(q)\big] 
\Imm\big[\Sigma(q)\big]\bigg| <1 \ ,
\end{equation}
for any $\shat < \shat_{max}$.  This 
then gives a conservative lower bound on $\Lambda_1$.
%assures $M_{max}>\sqrt{\shat_{max}$, and
%as shown in Table \ref{table:boundLambda1}.

Explicitly, we find, in the notation of \Eref{Sigmadef},
\begin{equation}\label{GSigmaCalc}
G_\OO^{(0)}(q)\Sigma(q)=
{1\over \Lambda^{2\Delta_1}}\langle \OO_1(q)\OO_1(-q)\rangle
%\int \frac{d^4k}{(2\pi)^4} 
\langle \ \sum_a G^a_\mn G^{a\mn}(q)\ \sum_b G^b_\mn G^{b\mn}(-q)\ \rangle
\end{equation} 
Keeping only the finite imaginary parts, our consistency condition becomes
\begin{equation}\label{GSigmaRelation}
\bigg|\Imm \big[G_\OO^{(0)}(q)\big] 
\Imm\big[\Sigma(q)\big]\bigg| 
 = 8 \times \frac{
  \sin(-\pi\Delta_1) \Gamma[2-\Delta_1]}{(4\pi)^{2\Delta-2}
  \Gamma[\Delta_1]} \times \frac{2}{\pi}
\left(\frac{q^2}{\Lambda_1^2}\right)^{ \Delta_1} < 1
%  \log\left[
%  \frac{q}{4 \pi \Lambda_1}\right]   \ ,
\end{equation}
for $q^2\leq{\shat_{max}}$.
Here the important prefactor of 8  counts the number of gluon states.
This condition in turn implies a lower bound on $\Lambda_1$.

%  To obtain a precise bound, we demand that the absolute value of
%\Eref{GSigmaRelation} be no larger than $\frac{1}{2}$ when
%also shown in Table \ref{table:boundLambda1}, as a range of scales,
%where the lower scale applies for $\Delta_2\to 1$ and the higher scale
%for $\Delta_2\to 2$.  

The uncertainties that arise here stem mainly from the ambiguity in
the criterion chosen.  For example, suppose we replaced $1$ on the
right-hand side of \Eref{GSigmaRelation} with $\frac{1}{2}$?
This would only change
$\Lambda_1$ by $(2)^{1/2\Delta_1}$, and strengthen our final bound
by exactly a factor of $\frac12$.  This is, again, smaller
than the uncertainty in our bound that arises from the ambiguity in
defining $\shat_{max}$. 

As a final comment, we note that an analogous argument applies for
many other standard model operators, including those with higher spin,
putting similar lower bounds on the scale $\Lambda$.  We are not aware
of this constraint being accounted for elsewhere in the literature.

\begin{table}[b]
\begin{tabular}[c]{|| c || c|c|c|c|c|c|c|c|c|c||}\hline
\ \ \ \ \ \ \ 
$\Delta_2$& 1.05 & 1.15 & 1.25 & 1.35 & 1.45 & 1.55 & 1.65 & 1.75 & 1.85 & 1.95 \\
$\Delta_1$ \ \ \  &&&&&&&&&& \\ \hline 
1.05 & {\bf 9.19} & {\bf 9.19} & {\bf 9.19} & {\bf 9.19} & 
{\bf 9.19} & {\bf 9.19} & {\bf 9.19} & {\bf 9.19} & {\bf 9.19} & {\bf 9.19} \\
1.15& {\bf 5.18} & {\bf 5.18} & {\bf 5.18} & {\bf 5.18} & 
{\bf 5.18} & {\bf 5.18} & {\bf 5.18} & {\bf 5.18} & {\bf 5.18} & {\bf 5.18} \\
1.25& {\bf 3.19} & {\bf 3.19} & {\bf 3.19} & {\bf 3.19} & 
{\bf 3.19} & {\bf 3.19} & {\bf 3.19} & {3.26} & {3.43} &  { 3.60} \\
%1.35& {\bf 2.41} & {\bf 2.41} & 2.49 & 2.70 & 2.89 & 3.08 & 3.26 &
% 3.43 & 3.59 & 3.75 \\
%1.45& {\bf 1.99} & 2.19 & 2.38 & 2.56 & 2.73 & 2.90 & 3.06 & 3.21 
%& 3.36 & 3.50 \\
%1.55& 1.89 & 2.07 & 2.23 & 2.40 & 2.55 & 2.70 & 2.84 & 2.98 & 3.11 & 3.23 \\
%1.65& 1.78 & 1.94 & 2.09 & 2.23 & 2.37 & 2.50 & 2.62 & 2.74 & 2.86 & 2.97 \\
%1.75& 1.68 & 1.82 & 1.95 & 2.07 & 2.19 & 2.31 & 2.42 & 2.53 & 2.63 & 2.72 \\
%1.85& 1.58 & 1.70 & 1.82 & 1.93 & 2.04 & 2.14 & 2.24 & 2.33 & 2.42 & 2.51 \\
%1.95& 1.49 & 1.60 & 1.70 & 1.80 & 1.89 & 1.98 & 2.07 & 2.15 & 2.23 & 2.31\\
1.35& {\bf 2.11} & {\bf 2.11} & 2.24 & 2.43 & 2.62 & 2.80 & 2.98 & 3.15 & 3.31 & 3.47 \\
1.45& 1.76 & 1.95 & 2.13 & 2.31 & 2.48 & 2.65 & 2.81 & 2.96 & 3.10 & 3.24 \\
1.55& 1.68 & 1.85 & 2.01 & 2.17 & 2.32 & 2.47 & 2.61 & 2.74 & 2.87 & 3.00 \\
1.65& 1.59 & 1.74 & 1.89 & 2.03 & 2.16 & 2.29 & 2.41 & 2.53 & 2.65 & 2.76 
\\
1.75& 1.50 & 1.64 & 1.77 & 1.89 & 2.01 & 2.12 & 2.23 & 2.34 & 2.44 & 2.54 \\
1.85& 1.42 & 1.54 & 1.65 & 1.76 & 1.87 & 1.97 & 2.07 & 2.16 & 2.25 & 2.34 
\\
1.95& 1.34 & 1.45 & 1.55 & 1.65 & 1.74 & 1.83 & 1.92 & 2.00 & 2.08 & 2.16 \\
\hline
\end{tabular}
\caption{The minimum values of $\Lambda_1$ (in TeV), as a function of
$\Delta_1$ and $\Delta_2$, allowed by the experimental constraints
from monojets and by the theoretical constraint that conformal
invariance be preserved below $\shat_{max}$ for the corresponding
$\Delta_1,\Delta_2$; 
see Table \ref{table:smax}.  Values constrained by monojet
data are shown in boldface.}
\label{table:boundLambda1}
\end{table}

\subsection{Summary of the bounds on $\Lambda_1$}

The bound we obtain from the more powerful of these two constraints,
as a function of $\Delta_1$ and $\Delta_2$, is shown in Table
\ref{table:boundLambda1}.  The constraint from jet-plus-MET
measurements at the Tevatron is most powerful at small $\Delta_1$,
while the constraint of conformal invariance is the dominant effect at
larger $\Delta_1$.  Notice that the conformal invariance constraints
give a bound that becomes stronger as $\Delta_2$ increases, for fixed
$\Delta_1$.  Note also that the bound
never dips below 1 TeV.  One should also keep in mind that bounds on
monojets at Fermilab are probably stronger now than those which are
currently published.  The published CDF study \cite{CDFjMET} relies
only on 1.1 pb$^{-1}$.  Though it is systematics-limited, it appears
that some of these systematic uncertainties are data
driven and will have decreased with higher
statistics.

\section{Bounds on $pp\to \fourgamma$ at the LHC}

 With the bounds on $\Lambda_1$ from Table \ref{table:boundLambda1}, we
may now obtain bounds on $\sigma(pp\to \fourgamma)$ using the condition
from earlier sections.  First we obtain bounds based on our central
values and naive tree-level results; then we discuss their
uncertainties.

\subsection{Bounds in the model of Feng, Rajaraman and Tu}

 Let us consider first the particular case studied in
\cite{FengRajTu}, where $\OO_1=\OO_2$, $\Delta_1=\Delta_2$ and
$\Lambda_1=\Lambda_2$.  
Because of the equal
$\Delta_i$, the processes $gg\to gggg$, $gg\to gg\gamma\gamma$, 
and $gg\to\fourgamma$ all have the same energy dependence, so unitarity
constrains their sum, generalizing \Eref{finalbound}:
\beq\label{manyfinalstates2}
\sigma(pp\to gggg;\shat)&+& \sigma(pp\to gg\gamma\gamma;\shat)
+\sigma(pp\to \gamma\gamma\gamma\gamma;\shat)\cr \cr
%\nonumber \cr % \nonumber \cr
%&\approx& 81\sigma(gg\to \gamma\gamma\gamma\gamma;\shat) 
&<&
\frac{N_0(\Delta_1)}{s}\
\left(\frac{{s}}{[\Lambda_1^{min}]^2}\right)^{\Delta_1}
\left(\frac{s}{\shat_{max}}\right)^{2\Delta_2}
%N_0(\Delta_1)\
%\left({\shat_{max}}\right)^{-2\Delta_2}(\Lambda_1^{min})^{-2\Delta_1}
% {s^{\Delta_1+2 \Delta_2-1}} 
\int_0^{1} d\tau\ \frac{dL_{gg}(\tau)}{d\tau}
{\tau^{\Delta_1+2 \Delta_2-1}} \ .
%\left({\shat_{max}}\right)^{-2\Delta_2}(\Lambda_1^{min})^{-2\Delta_1}
%N_0(\Delta_1) 
%\frac{1}{s}\int_0^{s} d\shat\ \frac{dL_{gg}(\shat/s)}{d(\shat/s)}
%{\shat^{\Delta_1+2 \Delta_2-1}}
\cr
&\ &\
\eeq
All processes listed here proceed through the hidden sector; QCD
contributions to $gg\to gggg$ are of course not to be included.

To go further, we use the fact that the amplitudes for these processes
are identical (since neither electromagnetic nor strong interactions
enter the calculation at leading order); one may view the calculation
as taking place in $U(3)$ instead of $SU(3)$-color, with the photon
being the ninth gluon.  The only non-trivial aspect is interference,
which could be precisely computed, but we will only estimate.

Label the gluons with an index $a=1,\dots 8$, with $a=9$ for the
photon.  Label the matrix element for $gg\to g^ag^ag^bg^b$ as ${\cal
M}_{ab}(k_1,k_2,k_3,k_4)$.  Only the sums $k_{ij}=k_i+k_j$ enter the
amplitude.  Then ${\cal M}_{ab} = F(k_{12}^2,k_{34}^2) +
F(k_{13}^2,k_{24}^2)\delta_{ab} + F(k_{14}^2,k_{23}^2)\delta_{ab}$.
Also for $a=b$ there is a reduction in phase space by 3, due to Bose
statistics.  The effect is that if the three terms in ${\cal M}_{aa}$ interfered
maximally throughout phase space (which they do not), 
we would have
\begin{equation}\label{manyfinalstates}
\sigma(gg\to gggg;\shat):
\sigma(gg\to gg\gamma\gamma;\shat):
\sigma(gg\to \gamma\gamma\gamma\gamma;\shat) = 80:16:3
\end{equation}
while with no interference the numbers above would be $64:16:1$.  Thus
the ratio of $\sigma(gg\to\fourgamma)$ to the total in \Eref{manyfinalstates2}
is 1/81 without
interference, while if interference is maximal everywhere in phase
space, the ratio is 1/33.  In most regions of phase space, one of the
three terms in ${\cal M}_{aa}$ will dominate, so interference
effects will be small.  But to be maximally
conservative, since we have not performed the computation, we take the
ratio 1/33 for our upper bound.  A full computation (or even a more
detailed argument using the power-law dependence of $F$) would
probably lead to a bound a factor of 1.5 to 2 stronger.

\begin{table}[tb]
\begin{tabular}[c]{|| c || c|c|c|c|c|c|c|c|c|c||}\hline
$\Delta_1=\Delta_2 $& 1.05 & 1.15 & 1.25 & 1.35 & 1.45 & 1.55 & 1.65 & 1.75 & 1.85 & 1.95 \\
\hline 
Max $\sigma$ [maximal interference] (in fb)& 
10.92& 19.26& 21.79& 15.63& 5.35& 1.98& 0.81& 0.34& 0.14& 0.07
%\hline
%Max $\sigma$ [no interference] (in fb)& 
%4.45& 7.85& 8.88& 6.37& 2.18& 0.81& 0.33& 0.14& 0.06& 0.03
% 1.46 &
% 2.77 &
% 3.27 &
% 2.74 &
% 0.94 &
% 0.35 &
% 0.14 &
% 0.06 &
% 0.02 &
% 0.01
\\
\hline
\end{tabular}
\caption{The maximum allowed values, {\it in femtobarns}, of the
cross-section for $pp\to\gamma\gamma\gamma\gamma$, as a function of
$\Delta_1=\Delta_2$, assuming $\OO_1=\OO_2$ and $\Lambda_1=\Lambda_2$,
as in \cite{FengRajTu}.  In this case --- see \Eref{manyfinalstates2}
--- both $pp\to gggg$ and $pp\to gg\gamma\gamma$ contribute to the
unitarity bound.  Since we have not performed the calculation
directly we simply assume maximal interference among diagrams; the 
true bound obtained from such a calculation would be stronger,
probably by a factor of 1.5--2.}
\label{table:boundO1isO2}
\end{table}

This gives bounds on $pp\to\fourgamma$ which are at least 33 times stronger
than obtained just from \Eref{finalbound}, reducing the allowable
4-photon cross-sections to less than 25 femtobarns, as shown in Table
\ref{table:boundO1isO2}.  
In particular, the case of $\Delta$ near 2, where the bound
in \cite{FengRajTu} was weakest, is where the unitarity bound is
the strongest, below 0.15 fb.

As we noted, this is obtained through a very conservative method,
assuming (contrary to fact) that interference is maximal everywhere.
Moreover, the reduction factor of 33 is increased to something closer
to 40 by QCD corrections and by including processes involving $Z$
bosons, such as $ggZZ, gg\gamma Z$, etc., in the final states.  It
would grow further if $\OO$ also couples to $SU(2)$ gauge bosons.  For
these reasons we view 10 fb as a more likely bound.  It is also
worth noting that, were
the bound saturated, requiring $\Delta\sim 1.2$ and
$\Lambda_1=\Lambda_1^{min}$ as given in Table
\ref{table:boundLambda1}, then jet-plus-MET signals would
significantly exceed Standard Model backgrounds at the LHC, giving 
a possible alternative discovery channel.

\subsection{General bounds}
The above situation is fairly generic.  There is no reason to expect
that any one process, especially one as experimentally attractive as
$gg\to\gamma\gamma\gamma\gamma$, dominates over all others.  However,
different processes cannot generically be combined together without
additional calculation.  For example, if $\OO_1\neq \OO_2$ and
$\Delta_1\neq\Delta_2$, as we
considered in most of this paper, then the choice of 
$s_{max}$ for $gg\to gggg$ is
not the same as for $gg\to\gamma\gamma\gamma\gamma$, 
and so their bounds are not simply related.
Furthermore, although the
four-gluon process is enhanced by color factors, it is proportional to
a different three-point coefficient; $C_{122}$ might be larger than
$C_{111}$, and the indeed the latter could even be zero.  In fact, we have
implicitly assumed $C_{111}=0$ in our main discussion, because a non-zero
value would give a stronger bound.

The strongest {\it model-independent} bound we can obtain
--- using the unitarity constraints we have discussed above --- is one
given by assuming that the {\it only} large process at the scale
$\shat_{max}$ is $gg\to\fourgamma$.  This is in principle
possible when $\OO_1\neq \OO_2$, so that $\Delta_2\neq
\Delta_1$ and $\Lambda_1\neq\Lambda_2$ in general.

Our bounds in this more general setting,
for various choices of $\Delta_1$ and $\Delta_2$, are shown
in Table \ref{table:boundsig4gam}.  
Interestingly, because our bounds
on $\Lambda_1$ are strong at low $\Delta_1$ but $\shat_{max}$ is
largest at higher $\Delta_1+2\Delta_2$, the bounds do not vary as
widely as a function of $\Delta_i$ as one might have imagined.  Note
that for those values of $\Delta_1,\Delta_2$ where the conformality
constraint is more important than the experimental bound from
jet-plus-MET, our bound depends only on $\Delta_1+2\Delta_2$; although
$\Lambda_1$ depends on $\Delta_1$ and $\Delta_2$ separately, the
conformality constraint and the total cross-section $\sigma(gg\to
\{X\})$ both depend on $\Lambda_1^{2\Delta_1}$, so that this
dependence cancels out of our limit.  

Our
bounds are smooth as the $\Delta_i$ pass through 2.  This is because
only the
imaginary part of the unparticle two-point functions arises in our
caculations.  As a result, none of our intermediate steps require
renormalization at $\Delta_i=2$.  
Conversely, note that we have
cut off our table at $\Delta_2=1.05$.  Although
our bound would formally
become still weaker as $\Delta_2\to 1$, there is 
a separate constraint in this region.  For $\Delta_2=1$, $\OO_2$ is 
a free field \cite{Mack}, satisfying the Klein-Gordon equation, and
therefore the OPE coefficient $C_{122}\to 0$ as $\Delta_2\to 1$ (with
the unique exception of the case where $\OO_1=(\OO_2)^2$, but then 
$C_{122}\to 1$ and $\Delta_1\to 2$ so the rate cannot be large.)
Consequently the four-photon
production cross-section generated through 
the three-point function $\langle\OO_1 \OO_2\OO_2 \rangle$ 
must be small as $\Delta_2\to 1$.

%\caption{The bound on the cross-section for $pp\to \fourgamma$.  Note that
%$\Lambda_1^{min}$ is a function of $\Delta_1$ only and is taken from
%Table \ref{table:boundLambda1}.  Values of $\shat_{max}$ are taken from 
%Table \ref{table:smax}.}

\begin{table}[b]
\begin{tabular}[c]{|| c || c|c|c|c|c|c|c|c|c|c||}\hline
\ \ \ \
$\Delta_2$& 1.05 & 1.15 & 1.25 & 1.35 & 1.45 & 1.55 & 1.65 & 1.75 & 1.85 & 1.95 \\
$\Delta_1$ &&&&&&&&&& \\ \hline 
% 1.05 & 120 & 56 & 28 & 14 & 8 & 4 & 2.4 & 1 & 1 & 0.5 \\
% 1.15 & 450 & 220 & 120 & 63 & 35 & 20 & 12 & 7 & 4 & 3 \\
% 1.25 & 950 & 490 & 270 & 150 & 84 & 49 & 30 & 18 & 11 & 7 \\
% 1.35 & 1740 & 940 & 480 & 230 & 110 & 55 & 29 & 16 & 9 & 5 \\
% 1.45 & 1690 & 720 & 320 & 150 & 77 & 40 & 21 & 12 & 7 & 4 \\
% 1.55 & 1090 & 480 & 220 & 110 & 55 & 29 & 16 & 9 & 5 & 3 \\
% 1.65 & 720 & 320 & 150 & 77 & 40 & 21 & 12 & 7 & 4 & 2 \\
% 1.75 & 480 & 220 & 110 & 55 & 29 & 16 & 9 & 5 & 3 & 2 \\
% 1.85 & 320 & 150 & 77 & 40 & 21 & 12 & 7 & 4 & 2 & 1 \\
% 1.95 & 220 & 110 & 55 & 29 & 16 & 9 & 5 & 3 & 2 & 1 \\
%\text {D1 D2 - $ > $} & 1.05 & 1.15 & 1.25 & 1.35 & 1.45 & 1.55 & \
%1.65 & 1.75 & 1.85 & 1.95 \\
%
1.05& 360& 170& 86& 45& 24& 13& 8& 5& 3& 2
%1.05 & 220 & 100 & 51 & 27 & 14 & 8 & 5 & 3 & 2 & 1
\\
1.15& 1270& 640& 330& 180& 100& 58& 34& 21& 13& 8 \\
1.25& 2530& 1320& 720& 400& 230& 138& 83& 49& 27& 15\\
1.35& 4270& 2330& 1120& 520 & 250 & 130 & 66 & 36 & 20 & 12 \\
%1.15 & 800 & 402 & 210 & 110 & 64 & 37 & 22 & 13 & 8 & 5 \\
%1.25 & 1640 & 860 & 470 & 260 & 150 & 89 & 54 & 33 & 21 & 14 \\
%1.35 & 2960 & 1620  & 910 

1.45 & 4020 & 1690 & 760 & 360 & 180 & 91 & 49 & 27 & 15 & 9 \\
1.55 & 2580 & 1120 & 520 & 250 & 126 & 66 & 36 & 20 & 12 & 7 \\
1.65 & 1690 & 760 & 360 & 180 & 91 & 49 & 27 & 15 & 9 & 5 \\
1.75 & 1120 &  520 & 250 & 126 & 66 & 36 & 20 & 12 & 7 & 4 \\
1.85 & 760 & 360 & 180 & 91 & 49 & 27 & 15 & 9 & 5 & 3\\
1.95 & 520 & 250 & 126 & 66 & 36 & 20 & 12 & 7 & 4 & 3\\
\hline
\end{tabular}
\caption{The maximum allowed values, {\it in femtobarns}, of the
cross-section for $pp\to\gamma\gamma\gamma\gamma$, as a function of
$\Delta_1$ and $\Delta_2$, assuming $\OO_1$ and $\OO_2$ are different
operators.  (See Table \ref{table:boundO1isO2} for the stronger bounds
that apply if $\OO_1=\OO_2$.)  Note that when the condition on
$\Lambda_1$ comes from the constraint of conformality, the bound
depends only $\Delta_1+2\Delta_2$.  }
\label{table:boundsig4gam}
\end{table}

Even though we are considering a much larger class of models, the
limits we obtain are much stronger than those quoted in
\cite{FengRajTu}, especially at high $\Delta_1,\Delta_2$.  
(For $\OO_1=\OO_2$, as in \cite{FengRajTu}, but
generalizing by allowing $\Lambda_1\neq\Lambda_2$,
the constraints are given along the diagonal, and
are always below 1 pb.)
However, we note that our bounds for $\Delta_1\sim 1.5, \Delta_2\sim
1$ -- were they saturated -- would still represent cross-sections of
considerable phenomenological interest.  One might have up to a few
hundred events in the first year of running at the LHC.  

It is worth noting that where the bounds for $pp\to\fourgamma$
lie well below 100 fb or so, this 
channel might not be the discovery channel.
For the values of $\Lambda_1$ shown in
Table \ref{table:boundLambda1}, and for $\Delta_1\lsim 1.4$,
the rate for jet-plus-MET at the LHC (for jet $p_T$ cuts of 250 GeV)
is generally in the few pb range. 
This is somewhat larger than the standard model rate.  
%
%TABLE??? 
%
Even though this measurement will be challenging in the early days of
a new hadron collider, with substantial systematic errors, such a
large excess in this channel might be convincing.  This means that
discovery of the new sector may well occur through the jet-plus-MET
channel.  In particular, this would almost certainly be the case in
the model of \cite{FengRajTu}, given the tight (yet conservative)
bounds in Table~\ref{table:boundO1isO2}.  For larger $\Lambda_1$ or
larger $\Delta_1$ the excess in jet-plus-MET may not be measurable,
but also the four-photon rate would be even further reduced.

Before concluding, we should re-emphasize the logic of our
argument.  Our claim is that if the cross-section for this process
exceeds our bound, then conformal invariance must be strongly violated,
which means that the universality of the ``unparticle'' dynamics is
lost, and the calculations of \cite{FengRajTu}, which assumed conformal
invariance, are not valid.  Instead, the production rate, and the
kinematic distribution, would become highly model dependent.

But we should hasten to add that large four-photon rates from a {\it
more general} hidden sector are still possible.  The
bounds in Table \ref{table:boundsig4gam} only constrain a conformally
invariant hidden sector.  A large four-photon signal could come from
other, non-``unparticle'' hidden sectors --- in particular from hidden
valleys, which might or might not be conformal at high energy, but
at low energy have strongly-broken conformal invariance and a mass gap. 
Examples of such theories are given in \cite{hvun,hvglue}.
Consequently, the four-photon experimental search channel, along with
other multi-particle search channels, is of considerable interest in
any case, and should be pursued model-independently.  However,
kinematic distributions will be very different from those in
\cite{FengRajTu,fourleptons,new4gamma}, and are highly
model-dependent.

\subsection{Uncertainties on the bounds}

Our bounds, as they are upper bounds, do not need to account for any
 experimental considerations, such as triggering rates, acceptance or
 efficiency, event selection cuts and the like, which can only reduce
 the number of events.  Indeed such considerations enter only in our
 determination of $\Lambda_1^{min}$ from existing experimental data.
 Because the $gg\to\fourgamma$ cross-section is largest at large $\hat
 s$, giving four photons which typically have momenta in the few
 hundred GeV range, neither triggering, efficiency or even geometric
 acceptance are likely to reduce significantly the number of observed
 events at the LHC.  This is especially true if a loose criterion
 (such as demanding only three of the four photons be observed) is
 applied in the analysis.

Still, our results have multiple sources of uncertainties.  For
example, we ignored K-factors which would have given us a stronger
bound on $\Lambda_1$, but which also would have given us a larger
cross-section for $gg\to\{X\}$ and therefore a weaker bound on $pp\to
\fourgamma$; these effects most likely cancel to a good approximation.
We also did not use the most updated parton distribution functions,
and in any case applied them only in a leading order approximation.
We neglected some experimental efficiencies in our extraction of
$\Lambda_1$, but were conservative in our use of the CDF data from
\cite{CDFjMET}.  We included only the largest jet-plus-MET process at
the Tevatron, worked only at leading order, and treated errors in the
CDF data using crude estimates of systematic and statistical errors.
Also we have used results from only 1.1 inverse fb; unpublished
limits have probably improved somewhat.

But the dominant source of uncertainty in our bound comes from our
choice of the parameter $\zeta$ defining $\shat_{max}$, and for this
reason it does not make sense for us to reduce the uncertainties
mentioned in the previous paragraph.  We chose to use $\zeta=\frac23$
in \Eref{definesmax1}.  Using $\zeta=\frac12$ could loosen our bounds
by a factor of order 3 -- 5.  On the other hand, such a choice puts
the peak cross-section right at the value of $\shat$ where the
unitarity bound is kicking in, which means that conformal invariance
is breaking down precisely where a prediction is most needed.  One
could also argue that $\zeta=\frac34$ is a better choice, which
would tighten the bounds by a factor of order 2.  In any
case, one must view this choice as one of taste. 
% We
%believe the criteria we have selected are very conservative. 
% However,
%it is impossible to set a precise bound because of this inherent
%ambiguity.  
But in addition we think it highly unlikely that a strict 
unitarity bound would be fully saturated in any physical
model.  It is much more probable that either conformal invariance will
break down below $\shat_{max}$, or that the pointlike interaction
between the two sectors will develop a form factor below
$\shat_{max}$.  Thus we expect that typically {\it a breakdown
of the methods of \cite{FengRajTu} occurs well below
the energy where the $gg\to
\fourgamma$ cross-section formally would exceed the $gg\to\{X\}$
cross-section.}  In this sense, we expect that our bounds, though
imprecise, are actually quite conservative.

\section{Comments on other multi-particle processes}
\label{sec:otherprocesses}

There are many other processes to which this type of unitarity bound
should be applied, each with its own features which we did not fully
explore here.  In particular, this type of bound is powerful whenever
the couplings between the two sectors are non-renormalizable, a
condition which ensures that a process such as $gg\to\fourgamma$ grows
with energy relative to $gg\to\{X\}$.  (Actually it is enough that the
couplings involving the final state particles, in our case $\OO_2F_\mn F^\mn$,
be non-renormalizable.)

An example where our bound would not be strong is in the process $q\bar
q\to\ell^+\ell^-\ell^+\ell^-$ through three {\it scalar} operators of
$\Delta\sim 1$, as considered in \cite{fourleptons}.  Here the
operator coupling the two sectors 
(after the Higgs gets an expectation value) has
dimension near four if the $\Delta_i$ are not far
above 1.  But conversely, as was demonstrated in \cite{fourleptons},
the lack of rapid growth at high energy also means there is no
suppression at low energy, and therefore Tevatron limits are very
strong.  Meanwhile, our arguments do apply if the
$\Delta_i$ are significantly larger than 1.

We argued in Sec.~\ref{subsec:Motivate}, however, that this case is
not physically reasonable anyway.  Large flavor-changing neutral
currents are essentially impossible to avoid if one couples a new
sector through chirality-flipping operators (as would be the case for
scalars) to light quarks and leptons.

The problem of flavor-changing currents would be alleviated in models
where the couplings to the quarks and leptons are weighted by mass, so
that no additional flavor dynamics is introduced.  In this case one
might consider $gg\to b\bar b b\bar b$ or $gg\to
\tau^+\tau^-\tau^+\tau^-$.  Here the bounds from our methods would be
weak. Fermilab production of this process would not be strongly
constrained in the case of $b\bar bb\bar b$.  However the trilepton
searches at Fermilab would significantly constrain the four-tau final
state.  Another possibility would involve $gg\to\gamma\gamma b\bar b$
or $ gg\to\gamma\gamma t\bar t$.  Our bound for the sum of these
processes is roughly 30 times weaker than for $gg\to\fourgamma$.
Backgrounds of course are larger too, but limits from Fermilab on
$\gamma\gamma b\bar b$ may be rather weak, and on $\gamma\gamma t\bar
t$ will be very limited because of kinematic constraints and low
statistics.  This case might merit additional exploration.

Another possibility involves couplings of standard model particles to
non-scalar operators in the conformal field theory.  In some cases the
couplings to light quarks and leptons would be chirality preserving and
need not introduce any new flavor dependence. Because unitarity
requires vector operators have dimension 3 or greater, and tensor
operators to have dimension 4 or greater, their couplings to the
standard model are always non-renormalizable.  Four-particle final
states generated through vector operators have growing cross-sections.
This means Tevatron bounds on processes such as $q\bar q\to
\ell^+\ell^-\ell^+\ell^-$ via vectors operators are weak, but
conversely our unitarity constraints are very
strong.

For example, one option with no fine-tuning
might involve the possibility of a three-point
function involving two pseudoscalar operators and a vector operator.  
Consider the process $gg\to \gamma\gamma \ell^+\ell^-$ which
would arise in a theory which has, in addition to the two couplings in
\Eref{somecouplings}, a third coupling
\begin{equation}\label{thirdcoupling}
{1\over\Lambda_2^{\Delta_3}}{\cal V}_\mu \sum_i \bar E_i \sigma^\mu E_i
\end{equation}
where $E_i$ is a left-handed charged antilepton $e^+,\mu^+,\tau^+$.
Because the vector operator ${\cal V}_\mu$ must have dimension
$\Delta_{\cal V}\geq 3$, the constraints obtained via our methods are
10--30 times stronger than those for $gg\to\fourgamma$, with the
maximum allowed cross-section being of order 100 fb.

\section{Conclusions and Outlook}

We considered an example of a multi-particle process mediated by a
hidden sector that is conformally invariant, along the lines of
\cite{FengRajTu}.  Conformal invariance makes the process predictable,
in a way that depends only on the dimensions of the
operators, up to an overall normalization.  We have shown that the
total cross-sections for such processes are strongly constrained by
requiring both conformal invariance and unitarity.  The constraint
is generally stronger when the products of standard-model and hidden-sector
operators that appear in the action have dimensions significantly
larger than 4.  This is because such non-renormalizable interactions
generate cross-sections that grow rapidly with energy, and will become
larger than the total hidden-sector production
cross-section at an energy that is of order
$\Lambda$, the scale of the coupling of the two sectors.  

In particular, we saw that, in the model suggested by
\cite{FengRajTu}, the process $gg\to \fourgamma$ is constrained to lie
below 25 fb.  Moreover, for operators with dimension $\Delta \lsim
1.5$, saturating this bound would require a scale $\Lambda$ so low
that the rate for jet-plus-MET would be larger, even at moderate $p_T$,
than the standard model rate.  For operators with $\Delta\gsim 1.5$,
the bound on $gg\to\fourgamma$ is below 3 fb.

However, relaxing the restrictive conditions in
\cite{FengRajTu} allowed us to raise the limits on the four-photon
cross-section, giving substantial LHC signals potentially as large as
a few pb.  But we emphasize that we believe that this is only the
beginning of the story.  More sophisticated constraints from unitarity
appear possible.  If so, the quantitative results obtained here will
be tightened further.  We hope to report on this, and clarify the
phenomenological situation, in a subsequent publication.

As we noted, our methods apply more widely.  Processes such as
$gg\to\gamma\gamma b \bar b$ with scalar operators coupling to heavy
flavor fermion-bilinears, which grow more slowly with energy than
$gg\to\fourgamma$, may be less constrained by unitarity, while
processes involving vector-operators, such as
$gg\to\gamma\gamma\ell^+\ell^-$, which grow more rapidly, are more
constrained.  However, experimental constraints from Fermilab are
stronger in the former case than the latter, precisely because
of this difference in energy dependence.

Our quantitative results do suffer from some ambiguities.  On the one hand,
we have been very conservative in our numbers.  We believe that
realistic limits are at least a factor of 2 or 3 stronger than we have
claimed.  Also, in real models the bounds that we obtained will rarely
be saturated, and even when they are, it is unlikely that the process
which saturates the bound will be the easiest to observe, as
$gg\to\fourgamma$ would be.  On the other hand, one could take an even
more conservative view regarding our definition of $\shat_{max}$, and get
bounds weaker by a factor of 3 or so.  However there is no way to
weaken our bounds by much more than this, except by giving up
conformal invariance, and with it the model-independent predictions of
the ``unparticle'' scenario.

\

The work of A.D. was supported by NSF grant PHY-0905383-ARRA; that of
M.J.S. was supported by NSF grant PHY-0904069
and by DOE grant DE-FG02-96ER40959.  We are grateful to the Aspen Center for
Physics for hospitality during a portion of this research.

\end{document}